\documentclass[a4paper, twoside, captions=nooneline, fleqn, parskip=half, appendixprefix, headsepline, draft=false, bibliography=totoc, abstracton, cleardoublepage=plain]{scrreprt}
\pdfoutput=1

\usepackage[ngerman,english]{babel} %Mehrsprachigkeit
\usepackage[T1]{fontenc} %Kodierungen
\usepackage[ansinew]{inputenc}
\usepackage{color}
\usepackage{amsmath} %besserer Mathesatz
\usepackage{amsfonts} %bestimmte Symbole
\usepackage{amssymb} %noch mehr bestimmte Symbole
\usepackage{graphicx} %grafiken einbinden
\usepackage{dsfont} %Einheitsmatrix und so
\usepackage{multirow} %table multirow support
\usepackage{booktabs} %tabellen
\usepackage[font=normalsize]{subfig} %subfloat command
\usepackage{lmodern} %Schönere Schrift
\usepackage{slashed} %Feynman slash notation
\usepackage{braket} %guess what
\usepackage{url} % \url support.
\usepackage[sort&compress]{natbib} % Expands bib capabilities. sort&compress deals with multiple citations in a nice way.
\usepackage{feynmp} %Feynman-Diagramme (für figures}
\usepackage[pdftex]{hyperref} % Schön verlinkte PDF-datei

\newcommand{\lgr}[1]{\mathcal{L}_\text{#1}}

\newcommand{\zvek}[2]{\begin{pmatrix} #1 \\ #2 \end{pmatrix}}

\newcommand{\zmat}[4]{\begin{pmatrix} #1 & #2\\ #3 & #4 \end{pmatrix}}
\newcommand{\dmat}[9]{\begin{pmatrix} #1 & #2 & #3\\ #4 & #5 & #6 \\ #7 & #8 & #9\end{pmatrix}}

\newtheorem{definition}{Definition}
\newtheorem{theorem}{Theorem}

\newenvironment{titleformat1}
	{\sffamily \huge \bfseries} {\mdseries \normalsize \rmfamily}
\newenvironment{titleformat2}
	{\sffamily \large \bfseries} {\mdseries \normalsize \rmfamily}
\newenvironment{titleformat3}
	{\sffamily \large} {\normalsize \rmfamily}
\newenvironment{titleformat4}
	{\sffamily} {\mdseries}

\makeatletter
\hypersetup{pdftitle=\@title, pdfauthor=\@author} %Sets pdf file information correctly
\makeatother

\bibpunct{[}{]}{;}{n}{}{,}

\addto\captionsenglish{}

\title{Monte-Carlo Event Generation for a Two-Higgs-Doublet Model with Maximal CP Symmetry}
\subtitle{BSc Thesis}
\author{Johann Brehmer}
\date{\today}

\begin{document}

\begin{titlepage}
\begin{center}

\begin{titleformat3}
	Department of Physics and Astronomy\\
	University of Heidelberg\\
\end{titleformat3}

\vspace{4cm}

\begin{titleformat1}
	Monte-Carlo Event Generation\\
	for a Two-Higgs-Doublet Model\\
	with Maximal CP Symmetry\\
\end{titleformat1}

\vfill

\begin{titleformat2}
	Bachelor Thesis in Physics\\
\end{titleformat2}

\vspace{0.1cm}

\begin{titleformat4}
	submitted by\\
\end{titleformat4}

\vspace{0.1cm}

\begin{titleformat2}
	Johann Brehmer \\
\end{titleformat2}

\vspace{0.1cm}

\begin{titleformat4}
	born in Bremen (Germany)\\
\end{titleformat4}

\vspace{0.1cm}

\begin{titleformat2}
	2012 \\
\end{titleformat2}

\cleardoublepage

\ \\

\vfill

This Bachelor Thesis has been carried out by Johann Brehmer at the\\
Kirchhoff Institute for Physics in Heidelberg under the supervision of\\
Professor Hans-Christian Schultz-Coulon.

\vspace{1cm}

Corrected version from 01 May, 2012.

\end{center}
\end{titlepage}

\cleardoublepage

\selectlanguage{english}
\section*{\abstractname}

In \citep{Maniatis_MCPMFamiliesMassHierarchy} a two-Higgs-doublet model with maximal symmetry under generalised CP transformations, the MCPM, has been proposed. The theory features a unique fermion mass spectrum which, although not describing nature precisely, provides a good approximation. It also predicts the existence of five Higgs bosons with a particular signature.

In this thesis I implemented the MCPM into the Monte-Carlo event generation package MadGraph, allowing the simulation of any MCPM tree-level process. The generated events are in a standardised format and can be used for further analysis with tools such as PYTHIA or GEANT, eventually leading to the comparison with experimental data and the exclusion or discovery of the theory. The implementation was successfully validated in different ways. It was then used for a first comparison of the MCPM signal events with the SM background and previous searches for new physics, hinting that the data expected at the LHC in the next years might provide exclusion limits or show signatures of this model.

\selectlanguage{ngerman}
\section*{\abstractname}

In \citep{Maniatis_MCPMFamiliesMassHierarchy} wurde ein Zwei-Higgs-Doublet-Modell mit maximaler Symmetrie unter generalisierten CP-Transformationen vorgeschlagen. Dieses sogenannte MCPM hat ein einzigartiges Massenspektrum, das die Natur zwar nicht exakt beschreibt, aber eine gute Näherung darstellt. Das Modell sagt die Existenz von fünf Higgs-Bosonen mit speziellen Eigenschaften voraus.

In dieser Arbeit habe ich das MCPM in die Monte-Carlo-Simulation MadGraph implementiert. Damit lassen sich Ereignisse für beliebige Tree-Level-Prozesse generieren, die dank standardisierter Schnittstellen mit vielen Programmen wie PYTHIA oder GEANT weiter verarbeitet und schließlich zum Ausschluss oder zur Entdeckung des Modells in Beschleuniger-Experimenten verwendet werden können. Die Implementierung wurde mit verschiedenen Methoden erfolgreich validiert. Anschließend wurde damit ein erster Vergleich einiger MCPM-Prozesse mit dem Untergrund aus Standardmodell-Prozessen sowie mit bisherigen experimentellen Studien durchgeführt. Die Ergebnisse deuten darauf hin, dass sich aus den Daten, die am LHC in den nächsten Jahren erwartet werden, entweder Ausschlussgrenzen oder Spuren des MCPM gewinnen lassen könnten.

\selectlanguage{english}

\cleardoublepage

\pagestyle{headings}

\tableofcontents

\cleardoublepage

\chapter{Introduction}
\label{ch:introduction}

The Standard Model of particle physics (SM) is one of the success stories of theoretical physics. This quantum field theory, which is based on a $SU(3)_C \times SU(2)_L \times U(1)_Y$ gauge group, describes all the fundamental particles that have been observed so far and the strong, electromagnetic and weak interactions extremely well. Only the Higgs boson it predicts is yet to be discovered.

However, from a theoretical point of view the SM is somewhat unsatisfactory. Its particle and mass structure seems arbitrary. There are three generations of fermions, but no explanation for this number. Of these fermions, the $t$, $b$ and $\tau$ are much heavier than the corresponding particles of the first and second generation\footnote{Here the $\overline{MS}$ quark masses evaluated at $\mu = v_0 \approx 246$ GeV have been used \citep[see][]{Maniatis_MCPMFamiliesMassHierarchy}.}:
\begin{align}
	\frac {m_e} {m_\tau} &\approx 3 \cdot 10^{-4} &\quad \frac {m_\mu} {m_\tau} &\approx 6 \cdot 10^{-2} \nonumber \\
	\frac {m_u} {m_t} &\approx 1 \cdot 10^{-5} &\quad \frac {m_c} {m_t} &\approx 4 \cdot 10^{-3} \nonumber \\
	\frac {m_d} {m_b} &\approx 1 \cdot 10^{-3} &\quad \frac {m_s} {m_b} &\approx 2 \cdot 10^{-2}
	\label{eq:MassHierarchyNature}
\end{align}
There is no explanation for this mass hierarchy, either. Finally, the SM features the Cabibbo-Kobayashi-Maskawa (CKM) matrix $V$. The absolute values of its entries are given by \citep{PDG}
\begin{equation}
	\dmat {\left|V_{11}\right|} {\left|V_{12}\right|} {\left|V_{13}\right|} {\left|V_{21}\right|} {\left|V_{22}\right|} {\left|V_{23}\right|} {\left|V_{31}\right|} {\left|V_{32}\right|} {\left|V_{33}\right|} \approx \dmat {0.97} {0.23} {0.00} {0.23} {0.97} {0.04} {0.01} {0.04} {1.00},
	\label{eq:CKMNature}
\end{equation}
which is to a good approximation unity. Again, the SM does not provide any reason for this. Not counting the representation assignments, there are 18 free parameters in the theory \citep{Nachtmann}.

A lot of research has been done on possible extensions of the SM. One ansatz is the extension of its scalar sector from one scalar doublet to two scalar doublets. By requiring that such a two-Higgs-doublet model (THDM) is invariant under so-called generalised CP transformations, a model with intriguing features can be constructed: the maximally CP-symmetric two-Higgs-doublet model (MCPM). It has been developed and studied by various authors in the last years \citep{Maniatis_CPViolationGeometry, Ferreira_THDMGeometry, Maniatis_THDMStability, Maniatis_MCPMFamiliesMassHierarchy, Maniatis_MCPMPhenomenology, Maniatis_MCPMPhenomenologyAddendum, Maniatis_MCPMPhenomenologyRadiativeEffects, Maniatis_MCPMParameterSpace}. In the MCPM, the fermions of the first two generations are massless, while the $t$, $b$ and $\tau$ get a mass through electroweak symmetry breaking \citep{Maniatis_MCPMFamiliesMassHierarchy}. The CKM matrix is the unity matrix. Although this is not precisely as observed, it is a good approximation to nature. % These general Two-Higgs-doublet models (THDM) have interesting CP transformation properties. 

The MCPM predicts the existence of five Higgs bosons. These couple to the fermions in a unique way. One Higgs boson behaves similar to the Higgs boson in the SM. However, four of the five Higgs particles couple only to the second generation fermions, although these are massless in the theory \citep{Maniatis_MCPMPhenomenology, Maniatis_MCPMPhenomenologyAddendum}. Thus the model has a phenomenology that is different from the SM and other extensions of the SM. In chapter \ref{ch:MCPM} the MCPM and its phenomology are developed and explained.

One obvious question remains unanswered: can this theory be either discovered or excluded at the Large Hadron Collider (LHC) at CERN? To answer this, the experimental data has to be confronted to Monte-Carlo simulation results. In this Bachelor thesis, I implemented such a simulation of the MCPM based on the event generation package MadGraph \citep{MadGraph5}. With this tool it is now easy to numerically calculate tree-level cross-sections for arbitrary processes and to generate events for further analysis. This constitutes the main part of this project and is documented in chapter \ref{ch:implementation}.

The MadGraph implementation of the MCPM was systematically validated in different ways. The results are presented in chapter \ref{ch:validation}. Using this new tool, the most relevant MCPM signatures were compared to the background from SM processes. It was also checked whether the results of the search for a $Z'$ boson are likely to have already excluded the MCPM or a part of its parameter space. This analysis is the topic of chapter \ref{ch:analysis}.

\chapter{The maximally CP-symmetric two-Higgs-doublet model (MCPM)}
\label{ch:MCPM}

In this chapter the maximally CP-symmetric two-Higgs-doublet model is developed. At first, the framework of two-Higgs doublet models is introduced. Generalised CP transformations are defined and it is analysed under which conditions a THDM is invariant under these transformations. Based on these criteria, the MCPM is constructed. Its particle spectrum and properties after electroweak symmetry breaking are derived. Finally, the phenomenology of the MCPM particles at colliders such as the LHC is sketched, including Higgs boson production mechanisms, their decay modes and the background from SM processes.

\section{General two-Higgs-doublet model}
\label{sec:THDM}

A general two-Higgs-doublet model is a theory based closely on the SM. It features the gauge group $SU(3)_C \times SU(2)_L \times U(1)_Y$ with gauge fields $G^i_\mu$, $W^i_\mu$ and $B_\mu$, corresponding to colour, weak isospin and weak hypercharge, respectively. There are also the usual fermions: for each generation $m$ there are up- and down-type quarks $u^m, d^m$ as well as charged leptons $e^m$ and neutrinos $\nu^m$. The theory is chiral, i.\,e.\ left- and right-handed fermion fields are treated differently. The left-handed fermions are organised in $SU(2)_L$ doublets
\begin{equation}
	q_{L}^m = {\zvek {u^m} {d^m}}_L \quad \text{and} \quad l_{L}^m = {\zvek {\nu^m} {e^m}}_L,
\end{equation}
while the right-handed fermions $u_{R}^m$, $d_{R}^m$, $e_{R}^m$ transform as singlets. Let us for now consider $n$ generations of these fermions (later the observed three families will be assumed).

The gauge and fermionic sectors of the Lagrangian are given by
\begin{align}
	\lgr{Gauge} &= - \frac 1 4 G_{\mu\nu}^i G^{i \mu\nu} - \frac 1 4 W^i_{\mu\nu} W^{i \mu \nu} - \frac 1 4 B_{\mu\nu} B^{\mu \nu} , \\
	\lgr{Fermions} &= \bar q_{L}^m i \slashed D q_{L}^m + \bar l_{L}^m i \slashed D l_{L}^m + \bar u_{R}^m i \slashed D u_{R}^m  + \bar d_{R}^m i \slashed D d_{R}^m + \bar e_{R}^m i \slashed D e_{R}^m
	\label{eq:THDMGaugeFermions}
\end{align}
following the usual conventions. A detailed explanation of the symbols and quantities used here can be found in appendix \ref{app:LagrangianBeforeSB}.

All of this is just as in the SM. However, the scalar sector is different. Instead of one complex scalar doublet, there are two such fields, denoted by
\begin{equation}
	\varphi_i(x) = \zvek {\varphi^+_i (x)} {\varphi^0_i (x)} \quad (i = 1,2).
\end{equation}
Both of these scalar doublets are assigned weak hypercharge\footnote{In the context of supersymmetric models, the two scalar doublets are usually assigned different weak hypercharges of $\frac 1 2$ and $- \frac 1 2$ respectively. However, this is only a matter of convention, see \citep{Maniatis_CPViolationGeometry} for details.} $y = \frac 1 2$. Their properties are given by the scalar sector of the Lagrangian density,
\begin{equation}
	\lgr{Scalar} = (D_\mu \varphi_i)^\dagger (D^\mu \varphi_i) - V(\varphi_1, \varphi_2).
	\label{eq:THDMScalar}
\end{equation}
$V(\varphi_1, \varphi_2)$ is the Higgs potential. Its most general gauge-invariant and renormalisable form is
\begin{equation}
	V(\varphi_1, \varphi_2) = a_{ij} \varphi_i^\dagger \varphi_j + b_{klmn} (\varphi_k^\dagger \varphi_l) (\varphi_m^\dagger \varphi_n)
	\label{eq:THDMPotential}
\end{equation}
with arbitrary coefficients $a_{ij}, b_{klmn} \in \mathbb{R}$.

The last part of the Lagrangian describing the THDM are the Yukawa terms coupling the fermionic to the scalar fields:
\begin{equation}
			\lgr{Yukawa} = -\bar{e}_{R}^m C^{imn}_{l} \varphi_i^{\dagger} l_{L}^n + \bar{u}_{R}^m C^{imn}_{u} \tilde{\varphi}_i^{\dagger} q_{L}^n - \bar{d}_{0 R}^m C^{imn}_d \varphi_i^{\dagger} q_{L}^n + \text{h.\,c.} 
			\label{eq:THDMYukawa}
\end{equation}
The Yukawa couplings $C_u^i, C_d^i, C_l^i$ are matrices in the generation space and 
	\[\tilde{\varphi}_i^\dagger = \varphi_i^T \zmat 0 1 {-1} 0. \]

Now all parts of the THDM can be put together.
\begin{definition}[THDM]
	A general two-Higgs-doublet model is a quantum field theory described by the Lagrangian density
	\begin{equation}
		\lgr{THDM} = \lgr{Gauge} + \lgr{Fermions} + \lgr{Scalar} + \lgr{Yukawa}
		\label{eq:THDMLagrangian}
	\end{equation}
	where the different sectors are given in \eqref{eq:THDMGaugeFermions}, \eqref{eq:THDMScalar} and \eqref{eq:THDMYukawa}.
\end{definition}

\section{Developing the MCPM}
\label{sec:DerivingMCPM}

\subsection{Generalised CP transformations}
\label{sec:GeneralisedCPTrafos}

In quantum field theories charge conjugation (C) and parity transformation (P) play important roles. Especially interesting is their combination. The CP transformation was long thought to be a fundamental symmetry of nature, until its violation was discovered in 1964 \citep{CroninFitch_CPVIolation}. It is strongly related to the question why there is more matter than antimatter in the universe, which has still not been answered satisfactorily \citep{Sakharov_MatterAntimatter}.

Under a standard CP transformation (denoted by $CP_s$) scalar fields transform as
\begin{equation}
	\varphi_i(x) \stackrel{CP_s}{\xrightarrow{\hspace*{0.8cm}}} \varphi_{i}^* (x')
\end{equation}
and spinor fields as
\begin{equation}
	\psi^m(x) \stackrel{CP_s}{\xrightarrow{\hspace*{0.8cm}}} i \gamma^0 \gamma^2 \psi^{m*} (x')
\end{equation}
with
\begin{equation}
	x = \zvek {x^0} {\mathbf{x}} \stackrel{CP_s}{\xrightarrow{\hspace*{0.8cm}}} x' = \zvek {x^0} {- \mathbf{x}}
	\label{eq:CoordinateTransformation}
\end{equation}
due to the parity transformation \citep[section 3.6]{PeskinSchroeder}\footnote{The fields may gain additional minus signs under a $CP_s$ transformation. This is a matter of convention.}.

In this standard CP transformation the spinor fields of the different generations do not mix, neither are the two Higgs fields transformed into each other. However, it is possible to define \emph{generalised CP transformations} that allow these kinds of mixings \citep{Ecker_GeneralisedCP}. For instance, such a transformation might consist of the usual charge conjugation and parity transformation, but additionally transform $\varphi_1$ into $\varphi_2$ and $\varphi_2$ into $\varphi_1$.

In other words, generalised CP transformation are standard CP transformations together with a change of basis in the scalar doublet space and a change of basis in the generation space. These basis changes are not totally arbitrary. Just like two successive standard CP transformations never change the physics of a system, two successive generalised CP transformations are not allowed to change the physical states of the particles either.

\begin{definition}[Generalised CP transformation]
A generalised CP transformation $CP_g$ transforms the spatial coordinates as in \eqref{eq:CoordinateTransformation}, scalar fields $\varphi_i$ ($i = 1,2$) as
\begin{equation}
	\varphi_i(x) \stackrel{CP_g}{\xrightarrow{\hspace*{0.8cm}}}  U_{\varphi\,ij} \varphi_j^{*} (x')
	\label{eq:GeneralisedCP1}
\end{equation}
and spinor fields $\psi^m$ with $n$ generations as
\begin{equation}
	\psi^m(x) \stackrel{CP_g}{\xrightarrow{\hspace*{0.8cm}}} i U_\psi^{mk} \gamma^0 \gamma^2 \psi^{k*} (x').
	\label{eq:GeneralisedCP2}
\end{equation}
The matrices $U_\varphi \in U(2)$, $U_\psi \in U(n)$ have to satisfy the condition that two successive gauge transformations do not change the fields at all or are at most equal to a gauge transformation:
\begin{equation}
	CP_g \circ CP_g \cong \mathds 1.
	\label{eq:GeneralisedCP3}
\end{equation}
The transformation properties of the gauge fields are described in \citep{Maniatis_MCPMFamiliesMassHierarchy}.
\end{definition}

Generalised CP transformations can be classified and further analysed, see for instance \citep{Maniatis_CPViolationGeometry}.

\subsection{Requirements for the MCPM}
\label{sec:MCPMRequirements}

It is now interesting to analyse which conditions a THDM has to satisfy to be invariant under $CP_g$ transformations. In \citep{Maniatis_MCPMFamiliesMassHierarchy} the concept of maximal CP symmetry has been introduced\footnote{Despite its name, maximal CP symmetry does not require a symmetry under \emph{all} generalised CP transformations. Requiring such a broad symmetry would necessarily lead to massless Higgs bosons. See \citep{Maniatis_MCPMFamiliesMassHierarchy} for more details.}\footnote{Note that in this thesis the abbreviation MCPM (for maximally CP-symmetric two-Higgs-doublet model) will exclusively refer to a specific model which will be defined later, not to any THDM with maximal CP symmetry.}:
\begin{definition}[Maximal CP symmetry]
A THDM is called maximally CP-\-sym\-me\-tric if for each $U_\varphi \in \left\{ \mathds 1, \sigma^1, i \sigma^2, \sigma^3 \right\}$ there is a suitable $U_\psi \in U(n)$ for each fermionic field such that the theory is invariant under the generalised CP transformation described by $U_\varphi$ and the different $U_\psi$. Here $\sigma^i$ are the Pauli matrices.
\end{definition}

It has been found \citep{Maniatis_MCPMFamiliesMassHierarchy} that this requirement is not enough to restrict the THDM to a theory with unambiguous mass spectrum and properties. Instead, it yields different possible scenarios. However, some of these cases predict a phenomenology that has already been ruled out by observations and can be excluded for this reason.

A THDM describing nature has to fulfill the following conditions \citep{Maniatis_MCPMFamiliesMassHierarchy}:
\begin{itemize}
	\item When the scalar fields acquire vacuum expectation values, the electroweak \linebreak[4] $SU(2)_L \linebreak[1] \times U(1)_Y$ symmetry is spontaneously broken to the electromagnetic $U(1)_Q$ symmetry. The Higgs potential is stable in the sense of \citep{Maniatis_THDMStability}.
	\item After symmetry breaking, the theory predicts none of the following:
	\begin{itemize}
		\item Large lepton flavour-changing neutral currents.
		\item Mass-degenerate massive fermions.
 		\item Massless Higgs bosons.
	\end{itemize}
\end{itemize}

\subsection{Consequences}
\label{sec:MCPMConsequences}

The requirement of maximal CP symmetry together with the phenomenological restrictions given above confines the Higgs potential \eqref{eq:THDMPotential} to a very specific form \citep{Maniatis_CPViolationGeometry, Maniatis_THDMStability}: 
\begin{theorem}
The Higgs potential of a maximally CP-symmetric THDM which satisfies the criteria given in section \ref{sec:MCPMRequirements} has the form
\begin{align}
			V(\varphi_1, \varphi_2) &= \xi_0 (\varphi_1^\dagger \varphi_1 + \varphi_2^\dagger \varphi_2) + \eta_{00} (\varphi_1^\dagger \varphi_1 + \varphi_2^\dagger \varphi_2)^2 + {\mu_1} (\varphi_1^\dagger \varphi_2 + \varphi_2^\dagger \varphi_1)^2 \nonumber \\
			&\quad +  {\mu_2} (-i \varphi_1^\dagger \varphi_2 + i \varphi_2^\dagger \varphi_1)^2 + \mu_3 (\varphi_1^\dagger \varphi_1 - \varphi_2^\dagger \varphi_2)^2
	\label{eq:MCPMPotential}
\end{align}
with real parameters satisfying
\begin{align}
		\label{eq:MCPMParameterConditions1}
	&\mu_1 \ge \mu_2 \ge \mu_3 ,\\
	&\eta_{00} > 0 ,\\ 
	&\mu_a + \eta_{00} > 0 \quad \text{for } a = 1,2,3 ,\\
	&\xi_0 < 0, \\
	&\mu_3 < 0.
		\label{eq:MCPMParameterConditions2}
\end{align}
\label{thm:MCPMPotential}
\end{theorem}

The consequences for the Yukawa couplings depend on the number of generations. The result for $n=1$ family of fermions is surprising \citep{Maniatis_MCPMFamiliesMassHierarchy}: 
\begin{theorem}
In a maximally CP-symmetric THDM with one generation that satisfies the criteria given in section \ref{sec:MCPMRequirements}, all Yukawa couplings $C_u^{imn}$, $C_d^{imn}$, $C_l^{imn}$ in \eqref{eq:THDMYukawa} must be zero.
\label{thm:MCPMYukawa1Gen}
\end{theorem}

In other words, in the maximally CP-symmetric THDM with only one fermion family no fermion will get a mass through electroweak symmetry breaking. This is remarkable: While in the SM the number of generations is totally arbitrary, the MCPM requires at least two generations for fermion masses to appear.

For $n=2$ generations the situation is different \citep{Maniatis_MCPMFamiliesMassHierarchy}:
\begin{theorem}
The Yukawa sector of a maximally CP-symmetric THDM with two generations (numbered 2 and 3 for reasons that will become clear later) which satisfies the criteria given in section \ref{sec:MCPMRequirements} is given by
\begin{align}
		\lgr{Yukawa} &= - c_l \left(\bar{e}^{3}_R \varphi_1^{\dagger} {l^3_L} - \bar{e}^2_R \varphi_2^{\dagger} l^2_L \right) \nonumber \\
		&\quad + c_u \left(\bar{u}^3_{R} \varphi_1^T i \sigma^2 q^3_L - \bar{u}^2_R \varphi_2^T i \sigma^2 q^2_L \right) \nonumber \\
		&\quad - c_d \left(\bar{d}^{\,3}_{R} \varphi_1^\dagger q^3_L - \bar{d}^{\,2}_R \varphi_2^{\dagger} q^2_L \right) + \text{h.\,c.}
		\label{eq:MCPMYukawa2Gen}
		\end{align}
	with real parameters $c_l, c_u, c_d \geq 0$.
\label{thm:MCPMYukawa2Gen}
\end{theorem}

Here $\varphi_1$ couples to one generation of fermions (labelled 3), $\varphi_2$ to the other (labelled 2), both with the same coupling constants. This will lead to a specific mass spectrum after symmetry breaking: one generation will become massive, while the other remains massless, as will be demonstrated in section \ref{sec:SymmetryBreaking}. So for two fermion families, the requirement of maximal CP symmetry yields an intrinsic mass hierarchy.

The general case of $n=3$ generations has not been fully analysed so far. However, a maximally CP-symmetric THDM with three generations, the so-called MCPM, can be created in a straightforward way: By identifying the solution for one generation with the first family ($u$, $d$, $e$, $\nu_e$) and the solution for two generations with the second ($c$, $s$, $\mu^-$, $\nu_\mu$) and third ($t$, $b$, $\tau$, $\nu_\tau$) families\footnote{Now the labels 2 and 3 for the two generations in \eqref{eq:MCPMYukawa2Gen} make sense.} and combining them, the MCPM is defined \citep{Maniatis_MCPMFamiliesMassHierarchy}. In other words, the three-generation MCPM is introduced with the following properties: The second and third generation of fermions couple to the second and first Higgs doublet respectively, and the first family of fermions is uncoupled to the scalar fields.
\begin{definition}[MCPM]
	The THDM with three fermion generations, Higgs potential \eqref{eq:MCPMPotential} and Yukawa sector
	\begin{align}
		\lgr{Yukawa} &= - c_l \left(\bar{e}^{3}_R \varphi_1^{\dagger} {l^3_L} - \bar{e}^2_R \varphi_2^{\dagger} l^2_L \right) \nonumber \\
		&\quad + c_u \left(\bar{u}^3_{R} \varphi_1^T i \sigma^2 q^3_L - \bar{u}^2_R \varphi_2^T i \sigma^2 q^2_L \right) \nonumber \\
		&\quad - c_d \left(\bar{d}^{\,3}_{R} \varphi_1^\dagger q^3_L - \bar{d}^{\,2}_R \varphi_2^{\dagger} q^2_L \right) + \text{h.\,c.}
		\label{eq:MCPMYukawa}
	\end{align}
	with parameters $\mu_1, \mu_2, \mu_3, \eta_{00}, \xi_0, c_l, c_u, c_d \in \mathbb{R}$ satisfying \eqref{eq:MCPMParameterConditions1} to \eqref{eq:MCPMParameterConditions2} and $c_l, c_u, c_d \geq 0$ is called MCPM. It is maximally CP-symmetric, as can be seen from theorem \ref{thm:MCPMYukawa1Gen} and theorem \ref{thm:MCPMYukawa2Gen}.
\end{definition}

This MCPM is the topic of this thesis. I should stress that it does not follow necessarily from the requirement of invariance under generalised CP transformations. It is not the most general maximally CP-symmetric THDM with three generations. However, it is a very manifest choice and leads to interesting properties after spontaneous symmetry breaking, as will be seen soon.

\subsection{Electroweak symmetry breaking}
\label{sec:SymmetryBreaking}

The Higgs potential \eqref{eq:MCPMPotential} is minimised by vacuum expectation values (vev) for the scalar doublets. Without loss of generality these values can be chosen such that only the second component of $\varphi_1$ gets a non-zero vev $v_0$ \citep{Maniatis_MCPMFamiliesMassHierarchy}: 
\begin{align}
	\braket{\varphi_1} &= \frac 1 {\sqrt{2}} \zvek 0 {v_0}, \\
	\braket{\varphi_2} &= \zvek 0 0,
\end{align}
where
\begin{equation}
	v_0 = \sqrt{\frac {- \xi_0} {\eta_{00} + \mu_3}}
\end{equation}
plays the same role as the vev in the SM.

The scalar doublets can then be expanded around this vacuum state. In unitary gauge this reads \citep{Maniatis_THDMStability}
\begin{align}
	\label{eq:SymmetryBreaking1}
	\varphi_1 (x) &= \frac 1 {\sqrt{2}} \zvek 0 {v_0 + \rho'(x)}, \\
	\varphi_2 (x) &=  \zvek {H^+(x)} {\frac 1 {\sqrt{2}} (h'(x) + i h''(x))}.
	\label{eq:SymmetryBreaking2}
\end{align}
Here $\rho'(x)$, $h'(x)$ and $h''(x)$ are real scalar fields, $H^+(x)$ is a complex scalar field with
\begin{equation}
	H^-(x) := \left(H^+(x)\right)^* .
\end{equation}
These fields are excitations from the vacuum state, so they represent physical particles: the physical Higgs bosons of the MCPM. The $\rho'$, $h'$ and $h''$ are uncharged and their own antiparticles, the $H^\pm$ pair turns out to be charged. As scalars all Higgs bosons have spin 0.

The gauge fields behave exactly as in the SM. The electroweak $SU(2)_L \times U(1)_Y$ symmetry is broken down to the electromagnetic $U(1)_Q$ symmetry. The $W^i_\mu$ and $B_\mu$ fields form the massive $W^\pm_\mu$ and $Z_\mu$ bosons and the massless photon $A_\mu$ \citep[see for instance][chapter 20.2]{PeskinSchroeder}.

In addition, the scalar sector \eqref{eq:THDMScalar} with Higgs potential \eqref{eq:MCPMPotential} yields mass terms for the Higgs bosons. These masses are given by \citep{Maniatis_MCPMFamiliesMassHierarchy}
\begin{align}
	\label{eq:MCPMHiggsBosonMasses1}
	m_{\rho'}^2 &= - 2 \xi_0, \\
	m_{h'}^2 &= 2 v_0^2 (\mu_1 - \mu_3), \\
	m_{h''}^2 &= 2 v_0^2 (\mu_2 - \mu_3), \\
	m_{H^\pm}^2 &= \frac {2 \mu_3 \xi_0} {\eta_{00} + \mu_3}
	\label{eq:MCPMHiggsBosonMasses2}
\end{align}
with 
\begin{equation}
	m_{h'}^2 \ge m_{h''}^2
	\label{eq:MCPMMassCondition}
\end{equation}
due to \eqref{eq:MCPMParameterConditions1}. The scalar sector also yields different interactions between the Higgs particles and gauge bosons.

In the Yukawa sector of the MCPM \eqref{eq:MCPMYukawa} only the fermions of the third generation couple to $\varphi_1$, and only $\varphi_1$ has a non-zero vacuum expectation value. Hence only the $t$, $b$ and $\tau$ become massive through electroweak symmetry breaking, while the fermions of the first and second generation stay massless. The tree-level fermion masses in the MCPM are \citep{Maniatis_MCPMFamiliesMassHierarchy}
\begin{align}
	\label{eq:MCPMFermionMasses1}
	m_t &= \frac {v_0 c_u} {\sqrt{2}}, \\
	m_b &= \frac {v_0 c_d} {\sqrt{2}}, \\
	m_\tau &= \frac {v_0 c_l} {\sqrt{2}}, \\
	m_f &= 0 \quad \text{for all other fermions } f.
	\label{eq:MCPMFermionMasses2}
\end{align}

\begin{figure}[htb]
	\center
	\vskip10pt
	\begin{fmffile}{HiggsFermionCoupling1}
		\begin{fmfgraph*}(50,50)
			\fmfleft{i}
			\fmfright{o1,o2}
			\fmflabel{$t, b, \tau$}{i}
			\fmflabel{$\rho'$}{o1}
			\fmflabel{$t, b, \tau$}{o2}
			\fmf{fermion}{i,v}
			\fmf{dashes}{v,o1}
			\fmf{fermion}{v,o2}
		\end{fmfgraph*}
	\end{fmffile}
	\hskip50pt
	\begin{fmffile}{HiggsFermionCoupling2}
		\begin{fmfgraph*}(50,50)
			\fmfleft{i}
			\fmfright{o1,o2}
			\fmflabel{$c, s, \mu$}{i}
			\fmflabel{$h', h''$}{o1}
			\fmflabel{$c, s, \mu$}{o2}
			\fmf{fermion}{i,v}
			\fmf{dashes}{v,o1}
			\fmf{fermion}{v,o2}
		\end{fmfgraph*}
	\end{fmffile}
	\hskip50pt
	\begin{fmffile}{HiggsFermionCoupling3}
		\begin{fmfgraph*}(50,50)
			\fmfleft{i}
			\fmfright{o1,o2}
			\fmflabel{$s, \mu^-$}{i}
			\fmflabel{$H^-$}{o1}
			\fmflabel{$c, \nu_\mu$}{o2}
			\fmf{fermion}{i,v}
			\fmf{dashes}{v,o1}
			\fmf{fermion}{v,o2}
		\end{fmfgraph*}
	\end{fmffile}
	\vskip10pt
	\caption{Couplings of Higgs bosons to fermions in the MCPM. While the $\rho'$ couples to the third generation, the other Higgs bosons only interact with the massless second-generation fermions.}
	\label{fig:MCPMHiggsFermionVertices}
\end{figure}
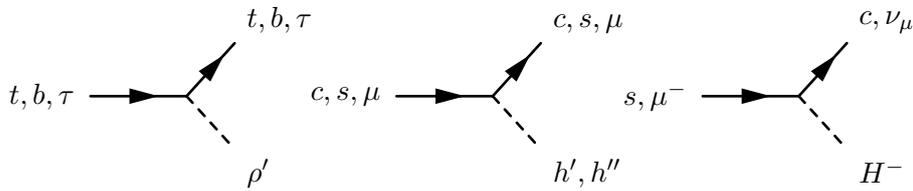

Similar to the SM Higgs boson, the $\rho'$ couples to all massive fermions, i.\,e.\ to the $t$, $b$ and $\tau$. However, the other Higgs particles behave differently. In \eqref{eq:MCPMYukawa} the second scalar doublet $\varphi_2$ couples only to the fermions of the second generation. Therefore its quantised excitations, the $h'$, $h''$ and $H^\pm$, couple to the $c$, $s$ and $\mu$, although these particles are massless in the MCPM. The Feynman diagrams of the Higgs-fermion couplings can be found in fig.\ \ref{fig:MCPMHiggsFermionVertices}. 

The strength of these interactions is even more surprising. Because the coupling of $\varphi_1$ to the third generation fermions and the coupling of $\varphi_2$ to the second generation fermions in \eqref{eq:MCPMYukawa} share the same coupling constants $c_l$, $c_u$ and $c_d$, the corresponding interaction vertices are weighted with the third-generation fermion masses. This is very different from the SM, where the second-generation fermions couple to the Higgs boson proportionally to the second-generation fermion masses, i.\,e. a lot more weakly. 

Finally, the CKM matrix turns out to be the unity matrix \citep{Maniatis_MCPMFamiliesMassHierarchy}, the quark mass eigenstates are weak interaction eigenstates as well. This again is not like in the SM, where the CKM matrix can be arbitrary (it is parametrised by four free parameters).

\section{The MCPM with three generations}
\label{sec:MCPM3Gen}

\subsection{Particle properties and couplings}
\label{sec:MCPMParticles}

Now the phenomenological properties of the MCPM can be put together. A list of all the particles and their key characteristics can be found in appendix \ref{app:MCPMParticles}. For the most important Feynman rules see appendix \ref{app:FeynmanRules}. There are four key differences between the MCPM and the SM:
\begin{enumerate}
	\item In the MCPM, the first two generations of fermions are massless at tree level. In the SM all fermions except the neutrinos are massive.
	\item In the MCPM, the CKM matrix is restricted to be unity. In the SM it is an arbitrary unitary matrix.
	\item There are five physical Higgs bosons (two of which are charged) in the MCPM compared to one in the SM.
	\item Four of these Higgs bosons couple to second-generation fermions, but weighted with the third-generation fermion masses. There is no equivalent structure in the SM.
\end{enumerate}

Looking at the first two points, the MCPM does not seem to describe nature very well. However, the theory gives quite a good approximation. While there is no doubt that there are massive fermions in the first two generations, they still are a lot lighter than the $t$, $b$ and $\tau$, see \eqref{eq:MassHierarchyNature}. The CKM matrix is certainly not quite $\mathds{1}$, but close to it, see \eqref{eq:CKMNature}. So the MCPM approximates nature not too badly with relatively few parameters (11 compared to 18 in the SM). Generalised CP transformations might be approximate symmetries of nature. They might even be exact symmetries that are broken in some way. Finally there is also the possibility that radiative corrections generate fermion masses for the first two generations. Thus although there are some obvious spots of bother, this theory should not be discarded too easily and can provide a starting point for more research on the CP transformation behaviour of our world.

So far the last two points of the list given above have neither been observed nor excluded by experiment. To discover or discard the MCPM as an exact, broken or approximate symmetry, they have to be studied carefully and checked at collider experiments like the LHC.

\subsection{Parameter space}
\label{sec:ParameterSpace}

There are 11 coefficients in the MCPM Lagrangian before symmetry breaking, not counting the representation assignments:
\begin{itemize}
	\item Gauge couplings: $g$, $g'$, $g_s$
	\item Higgs potential: $\mu_1$, $\mu_2$, $\mu_3$, $\xi_0$, $\eta_{00}$
	\item Yukawa coefficients: $c_u$, $c_d$, $c_l$
\end{itemize}
With \eqref{eq:MCPMFermionMasses1} to \eqref{eq:MCPMFermionMasses2} and \eqref{eq:MCPMHiggsBosonMasses1} to \eqref{eq:MCPMHiggsBosonMasses2}, these and all other relevant variables such as $v_0$ can be expressed through measurable quantities. A useful set consists of particle masses, the Fermi constant $G_F$ and the electromagnetic and strong coupling constants $\alpha$ and $\alpha_s$. The list of independent parameters of the MCPM then reads:
\begin{itemize}
	\item Couplings and gauge bosons: $\alpha$, $G_F$, $\alpha_s$, $m_Z$
	\item Fermion masses: $m_t$, $m_b$, $m_\tau$
	\item Higgs masses: $m_{\rho'}$, $m_{h'}$, $m_{h''}$, $m_{H^\pm}$
\end{itemize}
See appendix \ref{app:MCPMParameters} for a list of important quantities expressed through these independent parameters.

The gauge field properties and the fermion masses have already been measured extensively and with high precision, see \citep{PDG} for a summary. While there is no direct evidence for any of the Higgs particles so far, electroweak precision measurements can be used to exclude a part of the parameter space for the Higgs particle masses \citep{Maniatis_MCPMParameterSpace}. In addition, results from the search for the SM Higgs boson probably apply to the $\rho'$ as well, yielding strong restrictions on $m_{\rho'}$.

\section{Phenomenology at the LHC}
\label{sec:phenomenology}

\subsection{Higgs production}
\label{sec:HiggsProduction}

\begin{figure}[htb] 
	\center
	\vskip10pt
				\begin{fmffile}{QQFusion}
					\begin{fmfgraph*}(160,80)
						\fmfleft{i1,i2}
						\fmfright{o1,o3,o2}
						\fmflabel{$p$}{i1}
						\fmflabel{$p$}{i2}
						\fmflabel{$X$}{o1}
						\fmflabel{$X$}{o2}
						\fmflabel{$H$}{o3}
						\fmf{fermion,tension=1.5}{i1,v1}
						\fmf{fermion,tension=1.5}{i2,v2}
						\fmf{dbl_plain}{v1,o1}
						\fmf{dbl_plain}{v2,o2}
						\fmfv{decoration.shape=circle,decoration.filled=empty,decoration.size=10}{v1,v2}
						\fmf{fermion,label=$q$,tension=0.5}{v1,v3}
						\fmf{fermion,label=$\bar q'$,tension=0.5}{v3,v2}
						\fmf{dashes}{v3,o3}
					\end{fmfgraph*}
				\end{fmffile}
	\vskip10pt
	\caption{Production of Higgs bosons via quark-antiquark fusion. $X$ denotes the remnants of a proton, experimentally seen as jets.}
	\label{fig:QQFusionDiagram}
\end{figure}
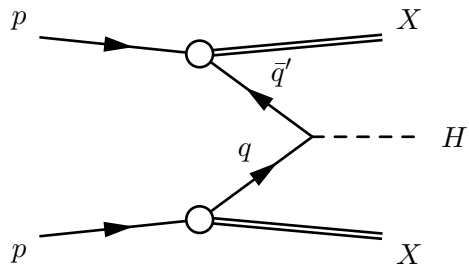

In proton-proton collisions, e.\,g.\ at the LHC, Higgs bosons can be produced via different processes. The most obvious one is quark-antiquark fusion, which is very similar to the Drell-Yan process \citep[section 17.4]{PeskinSchroeder}. The corresponding Feynman diagram is shown in fig.\ \ref{fig:QQFusionDiagram}. As discussed in section \ref{sec:MCPM3Gen}, the $\rho'$ couples only to $t$, $b$ and $\tau$, while the other Higgs bosons couple only to the second-generation fermions. Hence the following quark-level processes are possible (if energetically allowed):
\begin{align}
	t + \bar t &\rightarrow \rho', &\quad c + \bar c &\rightarrow h', &\quad c + \bar c &\rightarrow h'', &\quad  c + \bar s &\rightarrow H^+, \nonumber\\
	b + \bar b &\rightarrow \rho', &\quad s + \bar s &\rightarrow h', &\quad s + \bar s &\rightarrow h'', &\quad  s + \bar c &\rightarrow H^-. \nonumber
\end{align}

In \citep[section 4.1]{Maniatis_MCPMPhenomenology} the cross-sections for these kind of processes have been calculated as
\begin{align}
	\label{eq:HiggsProductionCrossSection1}
	\sigma (p p \rightarrow h' + X) \Big{|}_{q \bar q \text{ fusion}} &= \frac {\pi} {3 v_0^2 s} \left[ m_t^2 F_{c \bar c} \left( \frac {m_{h'}^2} s \right) + m_b^2 F_{s \bar s} \left( \frac {m_{h'}^2} s \right) \right], \\
	\sigma (p p \rightarrow h'' + X) \Big{|}_{q \bar q \text{ fusion}} &= \frac {\pi} {3 v_0^2 s} \left[ m_t^2 F_{c \bar c} \left( \frac {m_{h''}^2} s \right) + m_b^2 F_{s \bar s} \left( \frac {m_{h''}^2} s \right) \right], \\
	\sigma (p p \rightarrow H^+ + X) \Big{|}_{q \bar q \text{ fusion}} &= \frac {\pi} {3 v_0^2 s} \left(m_t^2 + m_b^2 \right) F_{c \bar s} \left( \frac {m_{H^\pm}^2} s \right), \\
	\sigma (p p \rightarrow H^- + X) \Big{|}_{q \bar q \text{ fusion}} &= \frac {\pi} {3 v_0^2 s} \left(m_t^2 + m_b^2 \right) F_{s \bar c} \left( \frac {m_{H^\pm}^2} s \right).
	\label{eq:HiggsProductionCrossSection2}
\end{align}
Here $\sqrt{s}$ is the center-of-mass energy and
\begin{equation}
	F_{q \bar q'} \left( \frac {m^2} s \right) = \int_0^1 \text d x_1\ \int_0^1 \text d x_2\ N_q(x_1) N_{\bar{q}'}(x_2) \delta \left( x_1 x_2 - \frac {m^2} s \right)
\end{equation}
where $N_q(x)$ is the parton distribution function (PDF) of the proton for quark $q$.

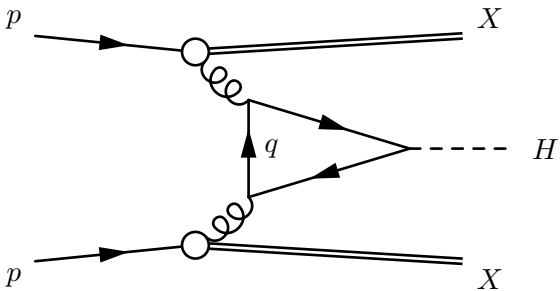
\begin{figure}[htb]
	\center
	\vskip10pt
				\begin{fmffile}{GGFusion}
					\begin{fmfgraph*}(200,100)
						\fmfleft{i1,i2}
						\fmfright{o1,o3,o2}
						\fmflabel{$p$}{i1}
						\fmflabel{$p$}{i2}
						\fmflabel{$X$}{o1}
						\fmflabel{$X$}{o2}
						\fmflabel{$H$}{o3}
						\fmf{fermion,tension=2}{i1,v1}
						\fmf{fermion,tension=2}{i2,v2}
						\fmf{dbl_plain}{v1,o1}
						\fmf{dbl_plain}{v2,o2}
						\fmfv{decoration.shape=circle,decoration.filled=empty,decoration.size=10}{v1,v2}
						\fmf{gluon}{v1,v4}
						\fmf{gluon}{v2,v5}
						\fmf{fermion,label=$q$,tension=0.33}{v4,v5}
						\fmf{fermion,tension=0.33}{v5,v3}
						\fmf{fermion,tension=0.33}{v3,v4}
						\fmf{dashes}{v3,o3}
					\end{fmfgraph*}
				\end{fmffile}
	\vskip10pt
	\caption{Production of Higgs bosons via gluon-gluon fusion.}
	\label{fig:GGFusionDiagram}
\end{figure}

Other production mechanisms include gluon-gluon fusion, see fig.\ \ref{fig:GGFusionDiagram}. For the SM Higgs boson this is the dominant production process \citep{Georgi_GluonFusion}. The $\rho'$ behaves in the same way. However, it turns out that for all the other MCPM Higgs bosons quark-antiquark fusion is dominant \citep{Maniatis_MCPMPhenomenology}. This difference in behaviour is caused by the fact that the $h'$, $h''$ and $H^\pm$ couple to the second-generation fermions, which are not that much suppressed in the PDFs, but the coupling strength is proportional to the heavy third-generation fermions. Hence quark-antiquark fusion is a lot stronger for these Higgs bosons than for the SM Higgs boson or the $\rho'$.

Still, the production cross-sections are rather small. Evaluating \eqref{eq:HiggsProductionCrossSection1} to \eqref{eq:HiggsProductionCrossSection2} at the LHC design energy $\sqrt{s} = 14$ TeV and for Higgs masses around 150 GeV yields cross-sections for each Higgs boson with an order of magnitude of 4000 pb (where the $H^\pm$ are slightly more likely to be produced than the $h'$ and $h''$). For heavier Higgs bosons around 400 GeV this value drops to mere 80 pb. For exact numbers see \citep[fig.\ 12]{Maniatis_MCPMPhenomenology} or appendix \ref{app:SystematicCheck}.

\subsection{Higgs decays}
\label{sec:HiggsDecays}

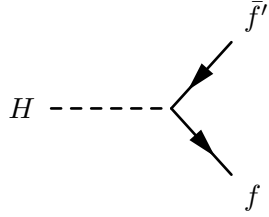
\begin{figure}[htb]
	\begin{center}
	\vskip10pt
				\begin{fmffile}{FermionicDecay}
					\begin{fmfgraph*}(90,60)
						\fmfleft{i}
						\fmfright{o1,o2}
						\fmflabel{$H$}{i}
						\fmflabel{$f$}{o1}
						\fmflabel{$\bar f'$}{o2}
						\fmf{dashes}{i,v}
						\fmf{fermion}{v,o1}
						\fmf{fermion}{o2,v}
					\end{fmfgraph*}
				\end{fmffile}
	\vskip10pt
	\end{center}
	\caption{Higgs boson decay into fermion-antifermion pair.}
	\label{fig:FermionicDecayDiagram}
\end{figure}

The Higgs bosons can decay in a number of ways. Similar to Higgs boson production, the most important modes stem from the fermion-fermion-Higgs vertices\footnote{Of course these decays are only possible if the Higgs bosons are heavy enough. Given current restrictions on the SM Higgs boson mass and taking into account that the $\rho'$ behaves in a very similar way, this is probably not true for the $\rho' \rightarrow t + \bar t$ mode.} \citep[section 3.1]{Maniatis_MCPMPhenomenology}:
\begin{align}
	\rho' &\rightarrow t + \bar t, &\quad h' &\rightarrow c + \bar c, &\quad h'' &\rightarrow c + \bar c, &\quad H^+ &\rightarrow c + \bar s, \nonumber\\
	\rho' &\rightarrow b + \bar b, &\quad h' &\rightarrow s + \bar s, &\quad h'' &\rightarrow s + \bar s, &\quad H^- &\rightarrow s + \bar c, \nonumber\\
	\rho' &\rightarrow \tau^- + \tau^+, &\quad h' &\rightarrow \mu^- + \mu^+, &\quad h'' &\rightarrow \mu^- + \mu^+, &\quad H^+ &\rightarrow \mu^+ + \nu_\mu, \nonumber \\
	& & & & & &\quad H^- &\rightarrow \mu^- + \bar{\nu}_\mu. \nonumber
\end{align}
The corresponding Feynman diagram is given in fig.\ \ref{fig:FermionicDecayDiagram}.

According to \citep{Maniatis_MCPMPhenomenology}, the decay width of such a process is given by
\begin{align}
	\Gamma (H \rightarrow f \bar f') &= \frac {N_c} {8 \pi v_0^2} \frac {\sqrt{m_H^4 + m_f^4 + m_{f'}^4 - 2 m_H^2 m_f^2 - 2 m_H^2 m_{f'}^2 - 2 m_f^2 m_{f'}^2}} {m_H} \nonumber \\
	&\times \theta \left( m_H - m_f - m_{f'} \right) \nonumber \\
	&\times \left[ |a|^2 + |b|^2 - \frac {(m_f + m_{f'})^2} {m_H^2} |a|^2 - \frac {(m_f - m_{f'})^2} {m_H^2} |b|^2 \right]
	\label{eq:HiggsDecayWidth}
\end{align}
where the vertex coefficients $a$, $b$ and the colour factor $N_c$ depend on the process and are given in table \ref{tbl:HiggsDecayCoefficients} in appendix \ref{app:DecayCoefficients}. These factors are proportional to the third-generation fermion masses. Because the $t$ is significantly heavier than the $b$ and $\tau$, the $t$ and $c$ channels dominate the decays, while the down-type quark and leptonic modes are suppressed by factors of roughly $10^{-3}$ to $10^{-5}$. The total decay width amounts to approximately 9 GeV for a Higgs mass of 150 GeV or 24 GeV for heavy Higgs bosons of 400 GeV.

In addition to these simple fermionic channels, Higgs bosons can decay into other Higgs bosons and gauge bosons. In \citep[section 3.2 to 3.4]{Maniatis_MCPMPhenomenology} it has been shown that these decays are irrelevant for light Higgs bosons, but become important for Higgs boson masses above roughly 400 GeV. However, analysing and simulating these channels goes beyond the scope of this project.

\subsection{Angular distribution and resonances}
\label{sec:MCPMDependencies}

Apart from total cross-sections, the dependency on angular variables such as the polar angle $\vartheta$, the pseudorapidity $\eta$ or the azimuthal angle $\varphi$ as well as the invariant mass of the decay products of the Higgs bosons is interesting.

The Higgs bosons in the MCPM are scalars. In their rest frame, there is no pre\-ferred spatial direction and their decay is isotropic. For reactions involving production and decay of Higgs bosons this implies
\begin{equation}
	\frac {\text d \sigma} {\text d \varphi^*} = \text{const}, \quad \frac {\text d \sigma} {\text d \cos \theta^*} = \text{const}\\
\end{equation}
where $\varphi^*$ and $\theta^*$ denote the angles in the rest frame of the virtual Higgs particle.

Note that this need not hold true in the lab frame. The quarks forming the Higgs bosons can each have arbitrary momenta\footnote{Up to an upper limit. This is described by the parton density functions.}, so the virtual Higgs particle need not be at rest in the lab frame. In fact, it is usually moving fast in either of the beam directions. Therefore the angular distribution of its decay products is distorted and peaks in the beam directions.

In processes including production and decay of an unstable particle such as a Higgs boson, the intermediate particle is not necessarily on-shell. Using the energies and momenta of its decay products, the invariant mass squared
\begin{equation}%
	p^2 = E^2 - \mathbf{p}^2
	\label{eq:InvariantMass}
\end{equation}
of a virtual Higgs boson can be reconstructed easily. The dependency of the cross-section on this invariant mass is described by the relativistic Breit-Wigner distribution \citep[section 2.4.4]{Berger}
\begin{equation}
	\frac {\text d \sigma} {\text d \sqrt{p^2}} = \frac k {(p^2 - m^2)^2 + m^2 \Gamma^2}.
	\label{eq:BreitWigner}
\end{equation}
Here $m$ and $\Gamma$ are the mass and decay width of the Higgs boson involved in the process, and $k$ is a constant of proportionality. This function peaks at $p^2 = m^2$, it describes the resonance shape of the process. Note that this is only an approximation valid in the limit of narrow widths, $\Gamma \ll m$. For broad peaks such as the Higgs processes of interest there will be some deviations. Most notably, the width of a broad resonance peak will slightly differ from the decay width of an on-shell intermediate particle \citep[see for instance][section 7.3]{PeskinSchroeder}.

\subsection{Standard model background}
\label{sec:SMBackground}

The MCPM Higgs bosons will never be produced on their own without any background; there are a lot of ordinary QCD and QED processes with the same final state that compete with Higgs production and decay.

\begin{figure}[htb]
	\center
	\vskip10pt
				\begin{fmffile}{BackgroundDrellYan}
					\begin{fmfgraph*}(220,80)
						\fmfleft{i1,i2}
						\fmfright{o1,o3,o4,o2}
						\fmflabel{$p$}{i1}
						\fmflabel{$p$}{i2}
						\fmflabel{$X$}{o1}
						\fmflabel{$X$}{o2}
						\fmflabel{$l^-$}{o3}
						\fmflabel{$l^+$}{o4}
						\fmf{fermion,tension=2.5}{i1,v1}
						\fmf{fermion,tension=2.5}{i2,v2}
						\fmf{dbl_plain}{v1,o1}
						\fmf{dbl_plain}{v2,o2}
						\fmfv{decoration.shape=circle,decoration.filled=empty,decoration.size=10}{v1,v2}
						\fmf{fermion,label=$q$,tension=0.5}{v1,v3}
						\fmf{fermion,label=$\bar q$,tension=0.5}{v3,v2}
						\fmf{boson,label=$\gamma^*,,Z$,tension=0.75}{v3,v4}
						\fmf{fermion,tension=0.5}{v4,o3}
						\fmf{fermion,tension=0.5}{o4,v4}
					\end{fmfgraph*}
				\end{fmffile}
	\vskip10pt
	\caption{Drell-Yan process. Quark and antiquark from colliding protons annihilate into a virtual photon or $Z$ boson, which decays into a pair of leptons.}
	\label{fig:BackgroundLeptonicDiagram}
\end{figure}
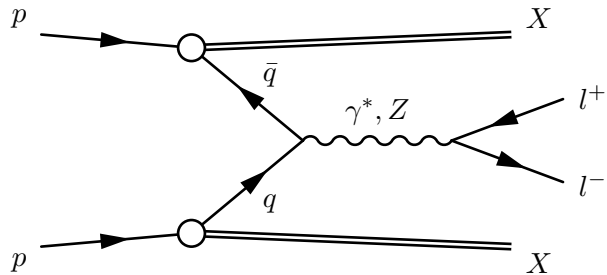

For the leptonic channels
\begin{align}
	p + p &\rightarrow \mu^- + \mu^+ + X, \nonumber \\
	p + p &\rightarrow \tau^- + \tau^+ + X, \nonumber
\end{align}
the most important SM background is the Drell-Yan process consisting of annihilation of a quark-antiquark pair into an excited photon or $Z$ boson, which decays into a pair of leptons \citep[section 17.4]{PeskinSchroeder}. The corresponding diagram is shown in fig.\ \ref{fig:BackgroundLeptonicDiagram}. There are similar diagrams for the $\mu^- \bar{\nu}_\mu$ and $\mu^+ \nu_\mu$ channels involving virtual $W^\pm$ bosons. All of these processes have large cross-sections. However, leptons can be detected efficiently, and it is easy to reconstruct the four-momentum and thus the invariant mass of the virtual particle in such a process.

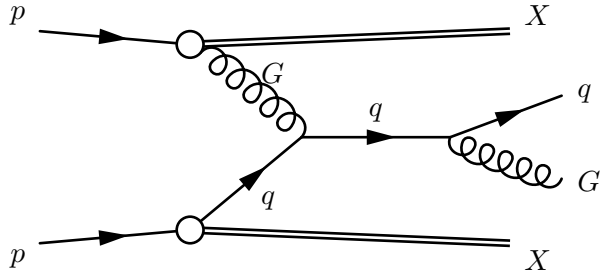
\begin{figure}[htb]
	\center
	\vskip10pt
				\begin{fmffile}{BackgroundQCD}
					\begin{fmfgraph*}(220,80)
						\fmfleft{i1,i2}
						\fmfright{o1,o3,o4,o2}
						\fmflabel{$p$}{i1}
						\fmflabel{$p$}{i2}
						\fmflabel{$X$}{o1}
						\fmflabel{$X$}{o2}
						\fmflabel{$G$}{o3}
						\fmflabel{$q$}{o4}
						\fmf{fermion,tension=2.5}{i1,v1}
						\fmf{fermion,tension=2.5}{i2,v2}
						\fmf{dbl_plain}{v1,o1}
						\fmf{dbl_plain}{v2,o2}
						\fmfv{decoration.shape=circle,decoration.filled=empty,decoration.size=10}{v1,v2}
						\fmf{fermion,label=$q$,tension=0.5}{v1,v3}
						\fmf{gluon,label=$G$,tension=0.5}{v3,v2}
						\fmf{fermion,label=$q$,tension=0.75}{v3,v4}
						\fmf{gluon,tension=0.5}{v4,o3}
						\fmf{fermion,tension=0.5}{v4,o4}
					\end{fmfgraph*}
				\end{fmffile}
	\vskip10pt
	\caption{Example for QCD background.}
	\label{fig:BackgroundQuarksDiagram}
\end{figure}

The quark channels are more complex. In addition to Drell-Yan-type processes, there are various QCD processes that yield two jets, see for instance fig.\ \ref{fig:BackgroundQuarksDiagram}. Cross-sections of the strong interaction are typically large. The situation gets even worse when taking hadronisation, jet tagging and reconstruction into account, which are complicated and therefore difficult to simulate and challenging experimentally.

So while the Higgs bosons of the MCPM decay mostly into quarks and much less into leptons, the huge background and the experimental difficulties of hadronic channels suggest that the leptonic channels are more promising \citep{Maniatis_MCPMPhenomenology}. Of these, the muonic events are most interesting, as the $h'$, $h''$ and $H^\pm$ couple to them in a way that is unique to the MCPM.

While there is certainly a large background hiding possible MCPM signals, the situation is not hopeless. The gauge bosons involved in the Drell-Yan-type processes have spin 1, while the Higgs bosons are scalars. This leads to a different angular distribution of the leptons or jets. In their rest frame, virtual photons and $Z$ bosons decaying to leptons approximately lead to a $(1 + \cos^2 \vartheta^*)$ distribution \citep[section 25.4]{Nachtmann}, which peaks in the beam directions. On the other hand, spin-0 particles decay isotropically as explained above. So by only looking at the central region of the detector, the SM background is reduced more strongly than the MCPM signal events. 

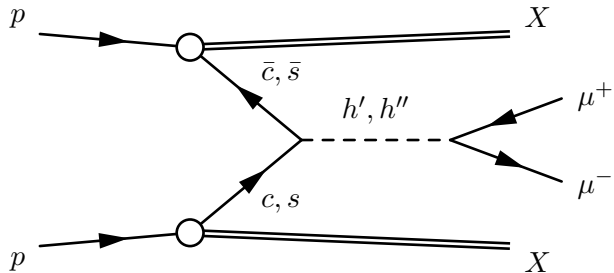
\begin{figure}[!bht]
	\center
	\vskip10pt
				\begin{fmffile}{FullReactionMuons}
					\begin{fmfgraph*}(220,80)
						\fmfleft{i1,i2}
						\fmfright{o1,o3,o4,o2}
						\fmflabel{$p$}{i1}
						\fmflabel{$p$}{i2}
						\fmflabel{$X$}{o1}
						\fmflabel{$X$}{o2}
						\fmflabel{$\mu^-$}{o3}
						\fmflabel{$\mu^+$}{o4}
						\fmf{fermion,tension=2.5}{i1,v1}
						\fmf{fermion,tension=2.5}{i2,v2}
						\fmf{dbl_plain}{v1,o1}
						\fmf{dbl_plain}{v2,o2}
						\fmfv{decoration.shape=circle,decoration.filled=empty,decoration.size=10}{v1,v2}
						\fmf{fermion,label=$c,,s$,tension=0.5}{v1,v3}
						\fmf{fermion,label=$\bar c,,\bar s$,tension=0.5}{v3,v2}
						\fmf{dashes,label=$h',,h''$,tension=0.75}{v3,v4}
						\fmf{fermion,tension=0.5}{v4,o3}
						\fmf{fermion,tension=0.5}{o4,v4}
					\end{fmfgraph*}
				\end{fmffile}
	\vskip10pt
	\caption{The full diagram describing production and muonic decay of the $h'$ and $h''$. There are similar diagrams involving $H^\pm$.}
	\label{fig:FullReactionMuonDiagram}
\end{figure}

In addition to separating the MCPM signature from the SM background, one also has to differentiate the MCPM signals from other SM extensions. Here the characteristic generation structure of the MCPM is important: While the $\mu^- \mu^+$ channels are relatively strong, the Higgs decays into other leptons are suppressed.

Taking everything into account, the MCPM has a unique phenomenology. The $\rho'$ behaves in a similar way as the SM Higgs boson and is therefore of less interest. For the $h'$, $h''$ and $H^\pm$, the main production channel at the LHC is predicted to be quark-antiquark fusion. These Higgs bosons decay mostly into $c$ and $\bar c$ quarks, but the resulting jets are hard to tag and reconstruct and there is a huge QCD background. Therefore the most promising channels are the muonic Higgs decays as shown in fig.\ \ref{fig:FullReactionMuonDiagram}, where the main background comes from Drell-Yan processes involving virtual photons or $Z$ bosons.

\chapter{Implementation of Monte-Carlo event generation}
\label{ch:implementation}

It is possible to calculate the production cross-section for Higgs bosons at the LHC using \eqref{eq:HiggsProductionCrossSection1} to \eqref{eq:HiggsProductionCrossSection2}. With \eqref{eq:HiggsDecayWidth} and similar expressions for other decay modes, the branching ratios for the different Higgs particles can be calculated. Assuming that the Higgs bosons are on-shell (i.\,e. "real" particles as opposed to virtual ones), the cross-section for the full process $p + p \rightarrow H + X \rightarrow f + \bar f + X$ is given by
\begin{equation}
	\sigma_{\text{full process}} = \sum_{H} \sigma_{H} \cdot B_{f \bar f}
	\label{eq:NarrowWidthApproximation}
\end{equation}
where $\sigma_H$ is the production cross-section for Higgs boson $H$ and $B_{f \bar f}$ is the branching ratio of the decay $H \rightarrow f + \bar f$. This is the narrow-width approximation and usually valid if the intermediate particle is much heavier than the initial and final particles \citep{Berdine_NWA}. This assumption is likely justified for the MCPM, which allows the calculation of cross-sections for the MCPM without any Monte-Carlo simulation, as has been done in \citep{Maniatis_MCPMPhenomenology}.

However in such a calculation a lot of effects cannot be easily taken into account, such as hadronisation and detector behaviour. For a full analysis a different approach is needed: the random generation of events with Monte-Carlo methods followed by the simulation of QCD and detector effects. This approach yields data that is directly comparable to experimental results.

The main part of this project was the implementation of such a Monte-Carlo event generation for the MCPM. The framework of MadGraph 5 has been used, which will be explained below. Afterwards the key features of the MCPM implementation are described.

\section{Introduction to MadGraph 5}
\label{sec:MadGraph}

MadGraph 5 \citep{MadGraph5} is the latest version of the matrix element generator MadGraph. It is bundled with the event generation package MadEvent. For simplicity, both will be referred to as MadGraph in the following. Given an arbitrary process for any implemented model, the program lists all contributing tree-level diagrams. For each of these diagrams, events are randomly produced and the cross-section or decay width is calculated.

The output of such a simulation is a list of events in the Les Houches Event File (LHEF) standard \citep{LHEF}. For each event, it includes the particles involved and their four-momenta. This can be read by programs such as PYTHIA \citep{PYTHIA}, with which hadronisation effects can be simulated. It is also possible to perform a detector simulation on the results, for instance using GEANT \citep{GEANT}. Event generation with MadGraph can thus be used as the first part of a complete Monte-Carlo analysis leading to a comparison of theory and experiment --- and ultimately the discovery or exclusion of new physics models such as the MCPM.

MadGraph is a very flexible framework and allows the implementation of new models in a simple way using the Universal FeynRules Output (UFO) format \citep{UFO}. Its major downside is the limitation to tree-level processes. However as seen in chapter \ref{sec:phenomenology}, the interesting and relevant processes do not involve any loops, so this drawback is acceptable.

\section{MCPM implementation}
\label{sec:MCImplementation}

In the course of this project, I have implemented the MCPM in MadGraph using the UFO interface \citep{UFO}. Basically, the model is described by three lists:
\begin{itemize}
  \item A list of the parameters of the MCPM. As independent parameters, the quantities listed in section \ref{sec:ParameterSpace} were used. For these variables, default values were chosen according to PDG recommendations \citep{PDG}. For the Higgs boson masses, values consistent with the restrictions presented in \citep{Maniatis_MCPMParameterSpace} were set.
  \item A list of the particles of the MCPM and their properties, including spin, mass, electromagnetic charge and colour charge.
  \item A list of the interaction vertices in the MCPM. For each vertex, the participating particles and the corresponding Feynman rule had to be given. Some important Feynman rules have already been calculated in \citep{Maniatis_MCPMPhenomenology}, others had to be derived first.
\end{itemize}
The MCPM model files are written in Python and span roughly 1300 lines. For more details, see appendix \ref{app:ImplementationParticlesParameters}.

All particles and couplings of the MCPM are included. This not only allows the analysis of signal processes involving Higgs bosons, but also the simulation of any tree-level background process. There is one important feature: strictly speaking, the MCPM forbids masses for the first two generations of fermions at tree-level. However nature seems to disagree with this prediction. This suggests that generalised CP transformations might only play the role of \emph{approximate} symmetries, as discussed in section \ref{sec:MCPM3Gen}. It may be sensible to analyse a model that combines the scalar sector of the MCPM with the experimental evidence for fermion masses and the CKM matrix \citep{Maniatis_MCPMPhenomenology}. For this reason the MadGraph implementation of the MCPM lets the user set fermion masses for the first two families as well. This affects phase space calculations, but the interactions remain unchanged. In a similar way the CKM matrix can be set to an arbitrary matrix, which changes SM processes, but not the Higgs boson vertices. With these parameters the user can decide whether to analyse the strict MCPM as described in chapter \ref{ch:MCPM} or a hybrid theory with fermion masses and a SM-like CKM matrix. If any SM processes play a role (for instance for background simulation), the latter seems to make more sense and will be used in the following.

Now the model can be used in the same way as other models are used in MadGraph. In order to generate events for a process, only a few commands are necessary. As an example, the lines
\begin{verbatim}
	import model MCPM -modelname
	generate p p > h1 > m- m+
	output -f
	launch
\end{verbatim}
lead to the generation of events for the process $p + p \rightarrow h' + X \rightarrow \mu^- + \mu^+ + X$. The free parameters can be set in a configuration file. For detailed instructions on how to use the model, see appendix \ref{app:ImplementationUsage}.

\chapter{Validation}
\label{ch:validation}

The MadGraph implementation of the MCPM was validated in two different ways. First, total cross-sections and decay widths were checked for a number of different scenarios. For this purpose Higgs-boson production and decay were treated independently. Second, angular dependencies and invariant mass distributions were analysed for full processes.

\section{Production cross-sections and decay widths}
\label{sec:SystematicCheck}

\subsection{Methodology}
\label{sec:SystematicCheckMethodology}

The Monte-Carlo simulation was validated by simulating different processes with MadGraph and comparing the resulting cross-sections as well as decay widths to the corresponding theoretical values. Production and decay were treated independently, i.\,e.\ the Higgs bosons were assumed to be on-shell.

As discussed in section \ref{sec:phenomenology}, the relevant production processes are
\begin{align}
	p + p &\rightarrow h' + X, \nonumber \\
	p + p &\rightarrow h'' + X, \nonumber \\
	p + p &\rightarrow H^- + X \nonumber
\end{align}
via quark-antiquark fusion. As before, $X$ denotes the fragmented remnants of the protons. The interesting decay modes include
\begin{align}
	h' &\rightarrow c + \bar c, &\quad h'' &\rightarrow c + \bar c, &\quad H^- &\rightarrow s + \bar c, \nonumber \\
	h' &\rightarrow \mu^- + \mu^+, &\quad h'' &\rightarrow \mu^- + \mu^+, &\quad H^- &\rightarrow \mu^- + \bar{\nu}_\mu. \nonumber
\end{align}
These three production and six decay processes were simulated.

For the MCPM parameters, PDG recommendations \citep{PDG} were used. Of course this leaves open the Higgs masses. Here four different points in parameter space have been chosen that fulfill two conditions:
\begin{itemize}
	\item They cover a wide range of the energies relevant at the LHC. In addition the ratios of the Higgs masses to each other are varied.
	\item All parameter sets agree with the electroweak precision measurements discussed in \citep{Maniatis_MCPMParameterSpace} within one standard deviation. They also agree with the stability criterium \eqref{eq:MCPMMassCondition}.
\end{itemize}
The chosen Higgs mass values are given in table\ \ref{tbl:SystematicCheckHiggsMasses}. 

\begin{table}[htb]
	\begin{tabular*}{\textwidth}{@{\extracolsep{\fill}} rrrrr}
		\toprule[0.12em]
		Parameter set & $m_{\rho'}$ [GeV] & $m_{h'}$ [GeV] & $m_{h''}$ [GeV] & $m_{H^\pm}$ [GeV]\\
		\midrule
		1 &	125	& 100	& 100 & 150	\\
		2	& 125	& 400	& 200 & 250 \\
		3	& 170	& 200	& 150 & 200 \\
		4	& 170	& 300	& 250 & 300 \\
		\bottomrule[0.12em]
 	\end{tabular*}
	\caption{Parameter sets used for the validation.}
	\label{tbl:SystematicCheckHiggsMasses}
\end{table}

The center-of-mass energy was set to $\sqrt s = 14 \text{ TeV}$ and no phase-space cuts have been applied. For the decays, 20,000 events have been produced in each MadGraph run. The production processes are more complex as they include multiple diagrams weighted by the PDFs of the proton. Here MadGraph was set to simulate 100,000 events in each run. The PDF CTEQ5L \citep{CTEQ5} was chosen for compatibility reasons.

The theoretical values were calculated for the same processes and settings. While the calculation of the decay widths \eqref{eq:HiggsDecayWidth} was straightforward and was done analytically, the production cross-sections \eqref{eq:HiggsProductionCrossSection1} to \eqref{eq:HiggsProductionCrossSection2} include the PDF and hence had to be calculated numerically, using the CTEQ5L Fortran libraries \citep{CTEQ5}.

For each of the nine processes given above and each of the four parameter sets in table \ref{tbl:SystematicCheckHiggsMasses}, a cross-section or decay width was simulated with MadGraph and a theoretical value was calculated. So in total there are 36 data points, each allowing the comparison of the Monte-Carlo result with the expectation.

\subsection{Results}
\label{sec:SystematicCheckResults}

The results of this procedure can be found in appendix \ref{app:SystematicCheck}. Fig.\ \ref{fig:SystematicCheckResiduals} shows residual plots, i.\,e.\ the deviations of MadGraph and theoretical results. Without a doubt, the Monte-Carlo results agree well with the expectation. For the decays, the numbers are exactly the same in all the five digits MadGraph indicates. The largest deviation in the production cross-section data is only 0.06 per cent, which is still well below one standard deviation of the Monte-Carlo results\footnote{This is even better than one might naively expect. Should there not be some statistical fluctuations leaving only 68\% of data points within one standard deviation of the theoretical values? The answer lies in the internal mechanisms of MadGraph \citep{MadGraph5}. The software calculates the cross-sections for individual quark-level processes precisely, the only uncertainty is introduced through the PDFs. However it gives out a standard deviation that does not represent this uncertainty of the calculated cross-section or decay width, but the magnitude of statistical fluctuations that are to be expected for an experiment observing the number of events set in MadGraph.}. The MCPM model in MadGraph passes its first test.

\begin{figure}[!h]
	\centering
	\subfloat[Higgs boson production]{\input{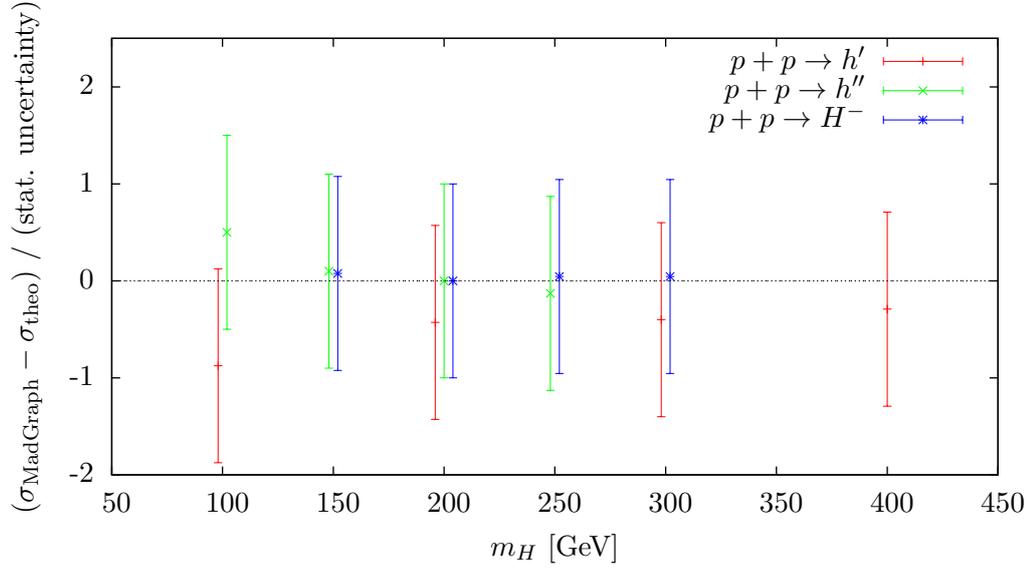} \label{fig:SystematicCheckProductionResiduals}}\\
	\subfloat[Higgs boson decay]{\input{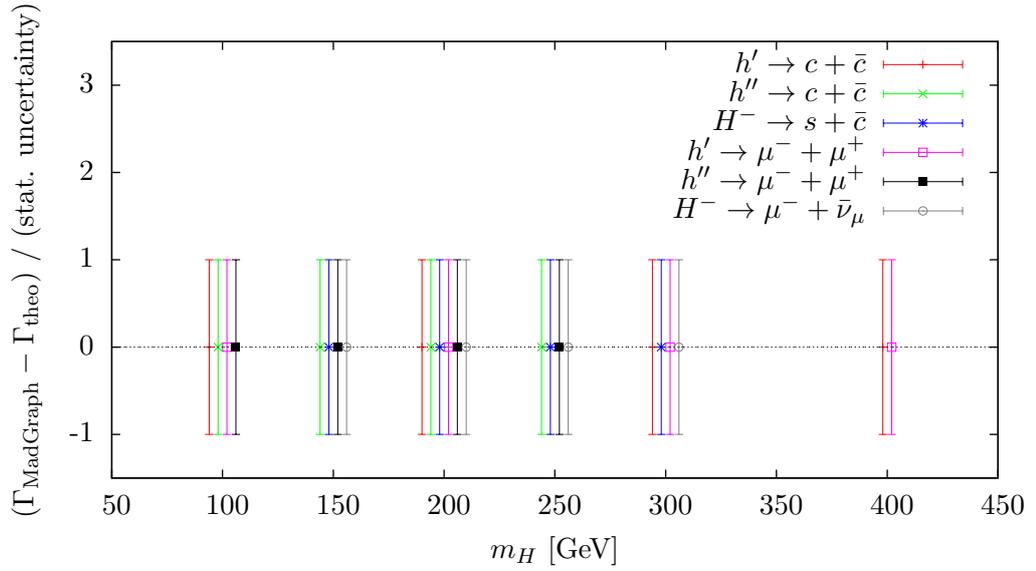} \label{fig:SystematicCheckDecayResiduals}}
	\caption{Residual plots showing the deviations of the MadGraph results and theoretical expectations. Good agreement is found for each process and parameter setting.}
	\label{fig:SystematicCheckResiduals}
\end{figure}

\section{Event distributions}
\label{sec:FullProcesses}

After validating the MadGraph results on the level of production cross-sections and decay widths, full processes including virtual Higgs bosons were simulated. The angular distributions and resonance shapes of the decay products were analysed.

\subsection{Methodology}
\label{sec:FullProcessesMethodology}

The processes 
\begin{align}
	\label{eq:FullAnalysisProcesses1}
	p + p &\rightarrow c + \bar c + X, \\
	p + p &\rightarrow s + \bar c + X, \\
	p + p &\rightarrow \mu^- + \mu^+ + X, \\
	p + p &\rightarrow \mu^- + \bar{\nu}_\mu + X
	\label{eq:FullAnalysisProcesses2}
\end{align}
of the form shown in fig.\ \ref{fig:FullReactionMuonDiagram} were simulated with MadGraph. Only MCPM Higgs bosons were allowed as intermediate particles, no SM background was included.

The parameters were chosen as in section \ref{sec:SystematicCheck} with three exceptions: every run included 100,000 events, and the newer PDF CTEQ6L \citep{CTEQ6} was used. The Higgs masses were set to
\begin{align}
	m_{\rho'}^A &= 125 \text{ GeV}, &\quad	m_{h'}^A &= 200 \text{ GeV}, \\
	m_{h''}^A &= 150 \text{ GeV}, &\quad 	m_{H^\pm}^A &= 150 \text{ GeV}.
	\label{eq:FullAnalysisHiggsMasses}
\end{align}
This parameter set will be referred to as set $A$ in the following. It complies with the restrictions found in \citep{Maniatis_MCPMParameterSpace}.

For these processes and parameters, MadGraph calculates the following total cross-sections:
\begin{align}
	\sigma_{c \bar c} &= 4980  \text{ pb}, \\
	\sigma_{s \bar c} &= 4760  \text{ pb}, \\
	\sigma_{\mu^- \mu^+} &= 0.175 \text{ pb}, \\
	\sigma_{\mu^- \bar{\nu}_\mu} &= 0.168  \text{ pb}
\end{align}
with negligible statistical errors.

In addition to calculating cross-sections, the resulting event files were examined using the tool MadAnalysis, which is included in the MadGraph 5 package. It allows the analysis of arbitrary distributions based on the four-momenta of the particles. In the following, angular dependencies and invariant mass distributions are shown. As MadAnalysis did not provide all the features needed, some routines had to be added by hand, including the correct handling of statistical uncertainties. ROOT \citep{ROOT} and GNUPlot \citep{GNUPlot} were used for fitting and plotting.

\subsection{Angular distribution}
\label{sec:AngularDistribution}

As described in section \ref{sec:MCPMDependencies}, the Higgs bosons should decay isotropically in their rest frame, which is the same as the center-of-mass frame of the decay products. The distribution is different in the lab frame. For the four Monte-Carlo samples generated, the angular distributions in both reference frames were analysed, some of the results are shown in fig.\ \ref{fig:PlotCosThetaMuMu} and fig.\ \ref{fig:PlotPhiMuMu}. More plots can be found in appendix \ref{app:FullProcesses}.

\begin{figure}[!h]
	\centering
	\subfloat[Distribution in the lab frame.]{\input{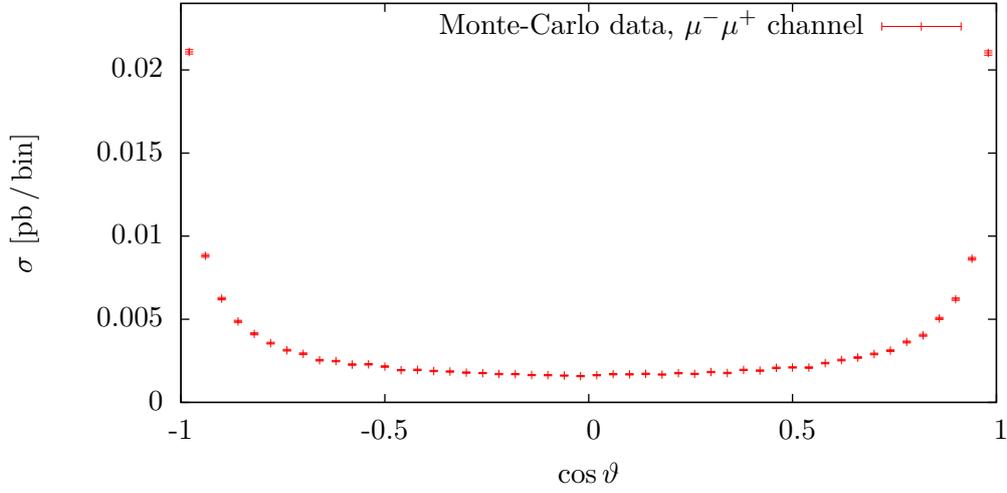} \label{fig:PlotLabCosThetaMuMu}}\\
	\subfloat[Distribution in the center-of-mass frame of the muons, which is the rest frame of the intermediate Higgs boson.]{\input{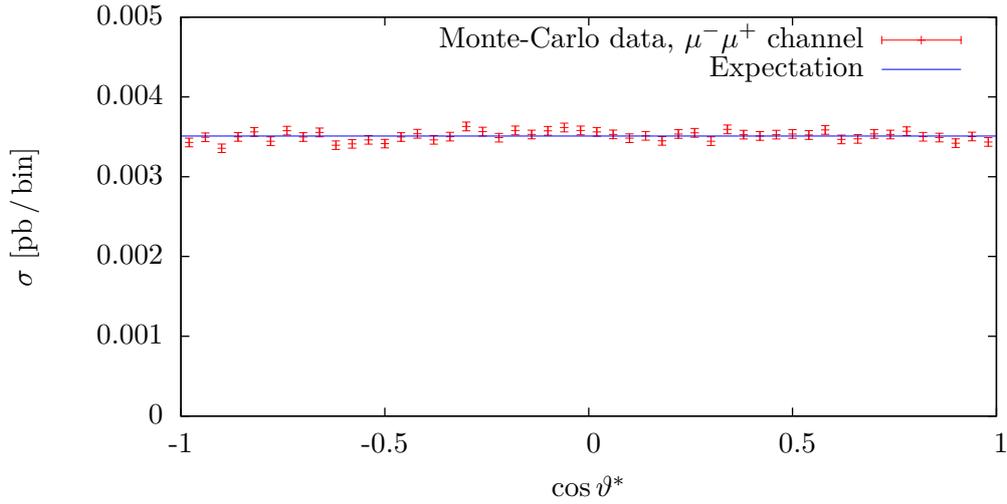} \label{fig:PlotComCosThetaMuMu}}
	\caption{Dependency of the $\mu^- \mu^+$ cross-section on the polar direction $\vartheta$ of the $\mu^-$. $\vartheta$ is defined to satisfy $\cos \vartheta = \pm 1$ in beam direction. $\cos \vartheta = 0$ represents a muon moving perpendicular to the beam axis. In the lab frame, the distribution is strongly peaked in beam direction due to the momentum of the virtual Higgs boson. After a boost into the center-of-mass frame of the muon pair, the distribution becomes isotropic.}
	\label{fig:PlotCosThetaMuMu}
\end{figure}

\begin{figure}[htb]
	\centering
	\input{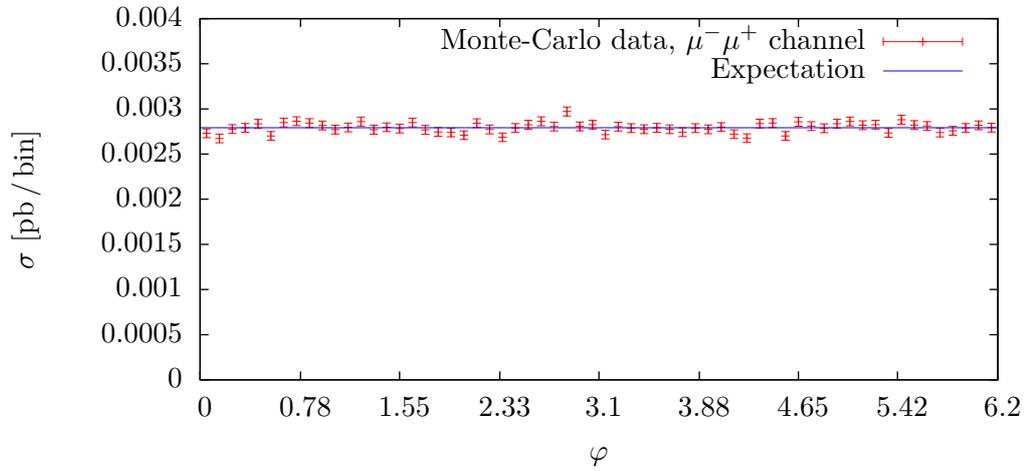}
	\caption{Dependency of the $\mu^- \mu^+$ cross-section on the azimuthal direction $\varphi$ of the $\mu^-$. To leading order, this distribution is the same in the lab frame and in the center-of-mass frame of the muons and should be constant. The Monte-Carlo data agrees very well with this expectation.}
	\label{fig:PlotPhiMuMu}
\end{figure}

For the $\cos \vartheta^*$ and $\varphi^*$ distributions in the Higgs boson rest frame, constant-value fits were performed. They agree well with the data, the $\chi^2$ values are listed in table \ref{tbl:AngularFitChiSquares}.

\begin{table}[htb]
		\centering
		\begin{tabular}{c r@{\,/\,}l r@{\,/\,}l}
			\toprule[0.12em]
			Channel & \multicolumn{2}{c}{$\cos \vartheta^*$ distribution} & \multicolumn{2}{c}{$\varphi^*$ distribution} \\
			& \quad \quad $\chi^2$ & $ndf$ & $\quad \quad \chi^2$ & $ndf$ \\
			\midrule
			$c \bar c$ & 40.7 & 49 & 53.8 & 61 \\
			$s \bar c$ & 53.3 & 49 & 58.2 & 61 \\
			$\mu^- \mu^+$ & 59.9 & 49 & 79.1 & 61 \\
			$\mu^- \bar{\nu}_\mu$ & 53.3 & 49 & 58.2 & 61 \\
			\bottomrule[0.12em]
		\end{tabular}
	\caption{Constant fits to the angular distributions in the Higgs boson rest frame.}
	\label{tbl:AngularFitChiSquares}
\end{table}

\subsection{Resonance shapes}
\label{sec:ResonanceShapes}

In addition to angular distributions, the dependency of the cross-section on the invariant mass of the Higgs boson is important. For each possible intermediate particle a resonance peak at its mass is expected. Its form should approximately be described by the relativistic Breit-Wigner distribution \eqref{eq:BreitWigner}, see section \ref{sec:MCPMDependencies}.

\begin{figure}[htb]
	\centering
	\input{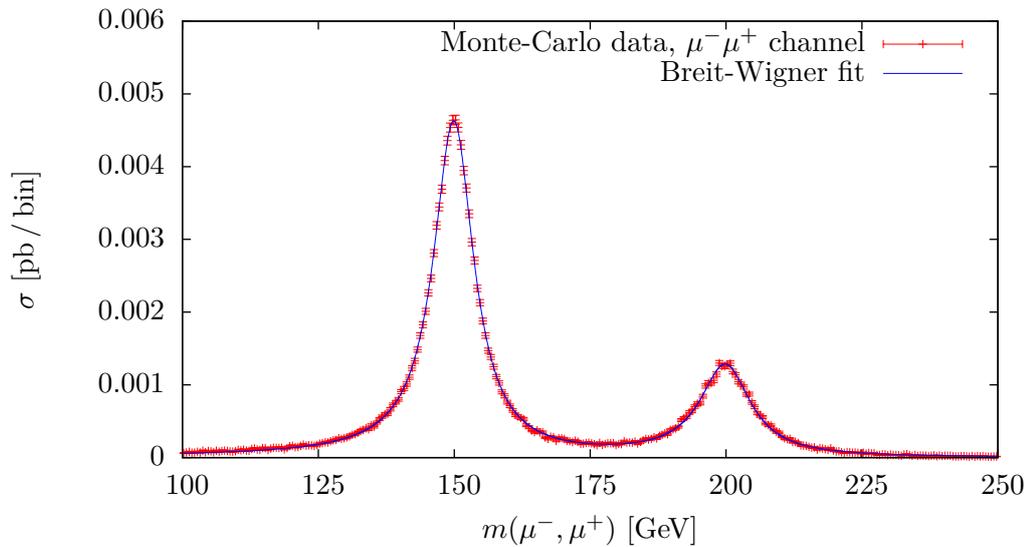}
	\caption{Dependency of the $\mu^- \mu^+$ cross-section on the invariant mass of the muon pair. This is identical to the invariant mass \eqref{eq:InvariantMass} of the virtual Higgs boson. There are resonances of the $h'$ at 200 GeV and of the $h''$ at 150 GeV. A superposition of two Breit-Wigner shapes \eqref{eq:BreitWigner} is fitted to the data with good agreement.}
	\label{fig:PlotResonanceShapeMuMu}
\end{figure}

For the four channels given above the invariant mass distribution was analysed. Fig.\ \ref{fig:PlotResonanceShapeMuMu} shows the resonance shape in the $\mu^- \mu^+$ channel, the others are similar. Breit-Wigner distributions were fitted to the data with good agreement. The resulting fit parameters were then compared to the Higgs boson masses \eqref{eq:FullAnalysisHiggsMasses} and the Higgs boson decay widths \eqref{eq:HiggsDecayWidth}, which were calculated for the dominant fermionic decay modes (see section \ref{sec:HiggsDecays}). The results can be found in table \ref{tbl:ResonanceShapeFitResults}.

\begin{table}[htb]
		\centering
		\begin{tabular}{c    r@{\,$\pm$\,}l r    r@{\,$\pm$\,}l r     r@{\,/\,}l}
			\toprule[0.12em]
			Channel & \multicolumn{3}{c}{$m$ [GeV]} & \multicolumn{3}{c}{$\Gamma$ [GeV]} & $\chi^2$ & $ndf$ \\
			& \multicolumn{2}{c}{Fit} & Theory & \multicolumn{2}{c}{Fit} & Theory & \multicolumn{2}{c}{ } \\
			\midrule
			$c \bar c$ &  149.90 & 0.02 & 150.00 & 9.00 & 0.04 & 8.83 & 515.5 & 294 \\
			& 199.87 & 0.04 & 200.00 & 11.24 & 0.08 & 11.77 & \multicolumn{2}{c}{ } \\
			\midrule
			$s \bar c$ & 149.94 & 0.02 & 150.00 & 8.81 & 0.03 & 8.83 & 726.3  & 297 \\
			\midrule
			$\mu^- \mu^+$ & 149.90 & 0.02 & 150.00 & 9.02 & 0.04 & 8.83 & 464.6 & 294 \\
			& 199.85 & 0.04 & 200.00 & 11.29 & 0.08 & 11.77 & \multicolumn{2}{c}{ } \\
			\midrule
			$\mu^- \bar{\nu}_\mu$ & 149.91 & 0.01 & 150.00 & 8.76 & 0.03 & 8.83 & 699.5 & 297 \\
			\bottomrule[0.12em]
		\end{tabular}
	\caption{Breit-Wigner fits to the resonance shapes.}
	\label{tbl:ResonanceShapeFitResults}
\end{table}

There are some small deviations. The reduced $\chi^2$ values are in the region of 2, and the fitted parameters slightly differ from the Higgs boson masses and decay widths. As explained in section \ref{sec:MCPMDependencies}, this is not surprising and a consequence of the relatively large widths of the Higgs bosons\footnote{Considering that the bin size used for this analysis is 0.5 GeV, binning effects also seem possible. However decreasing the bin size does not change the results by a relevant amount.}.

Taking everything into account, the MCPM implementation produces exactly the expected angular distributions and resonance peaks. Together with the cross-section check in section \ref{sec:SystematicCheck}, this result concludes its validation. Due to time constraints, not every process and not every parameter configuration have been tested. Nevertheless, I consider the model to be validated satisfactorily in order to be used for analysis.

\chapter{Analysis of MCPM signatures}
\label{ch:analysis}

The MCPM MadGraph implementation can now be used to analyse an important question: can the MCPM be discovered or excluded at the LHC? A full answer requires a thorough analysis including the simulation of fragmentation and detector behaviour, which goes well beyond the scope of this thesis. For the hadronic channels any superficial analysis using only quark-level events is meaningless, as hadronisation and jet reconstruction are complicated and important. These channels will not be further analysed.

However there is no hadronisation for leptons, and muons can be reconstructed relatively precisely. For the $\mu^- \mu^+$ channel the hard-process events generated by MadGraph are worth a look even without fragmentation and detector simulation. Using the new MadGraph implementation, the MCPM signatures are first compared to the SM background, both without and with phase-space cuts. Then ATLAS results from 2011 are checked against the MCPM predictions.

\section{Comparison of MCPM signals to SM background}
\label{sec:ComparisonSignalBackground}

In this section the MCPM signal events are compared to the SM background for the $\mu^- \mu^+$ channel. First the raw signals are analysed. Then basic $\eta$ and $p_T$ cuts motivated by detector properties are applied.

\subsection{Raw signal}
\label{sec:ComparisonRaw}

For this analysis, both signal and background events were generated with MadGraph\footnote{Signal and background can be treated separately. Because of the different chirality structures, there is no interference between signal and background diagrams.}. The relevant signal processes are once again the production of $h'$ and $h''$ bosons via quark-antiquark fusion with subsequent decay into $\mu^- \mu^+$ pairs, see fig.\ \ref{fig:FullReactionMuonDiagram}. As described in section \ref{sec:SMBackground}, the background is dominated by the $\gamma^*$ and $Z$ Drell-Yan processes. These lead to the same final state as the signal procesess and thus cannot be reduced easily.

The same parameters as in section \ref{sec:FullProcesses} were used. This includes the Higgs boson mass set $A$ \eqref{eq:FullAnalysisHiggsMasses} with $m^A_{h'} = 200$ GeV and $m^A_{h''} = 150$ GeV. In order to analyse a broader spectrum of resonances, a second Higgs mass set $B$ was chosen in addition:
\begin{align}
	m^B_{\rho'} &= 125 \text{ GeV}, &\quad	m^B_{h'} &= 400 \text{ GeV}, \\
	m^B_{h''} &= 250 \text{ GeV}, &\quad 	m^B_{H^\pm} &= 300 \text{ GeV}.
	\label{eq:ComparisonHiggsMasses}
\end{align}
For the background simulation, more than 500,000 events were generated in each run. All simulations were performed for the LHC design energy $\sqrt{s} = 14$ TeV as well as its current center-of-mass energy $\sqrt{s} = 7$ TeV. In the following the plots for 14 TeV are shown, the corresponding 7 TeV plots can be found in appendix \ref{app:SignalBackgroundComparison}.

Fig. \ref{fig:PlotComparisonBackgroundNoCuts14} shows the signal and background cross-sections. The resonance peaks of the Higgs bosons are tiny in comparison to the background. This can be quantified by comparing the MCPM and SM cross-sections for the invariant mass range\footnote{By choosing a smaller invariant mass bin, e.\,g.\ $m_H \pm \frac 1 2 \Gamma_H$, the ratio $\frac {\sigma_{\text{signal}}} {\sigma_{\text{background}}}$ increases. But from an experimental point of view it makes sense to use relatively large mass bins in order to get more statistics and due to detector resolution.} between $m_H - \Gamma_H$ and $m_H + \Gamma_H$. One finds
\begin{equation}
	\frac {\sigma_{\text{signal}}} {\sigma_{\text{background}}} \,\Big{|}_{\text{resonance}} \approx
	\begin{cases}
		3 \cdot 10^{-2} \quad \text{for } \sqrt{s} = 14 \text{ TeV} \\
		2 \cdot 10^{-2} \quad \text{for } \sqrt{s} = 7 \text{ TeV} \\
	\end{cases}
	\label{eq:CrossSectionRatioNoCuts}
\end{equation}
if the Higgs bosons are in the range of 150 to 250 GeV. For heavier Higgs bosons these factors decrease. See appendix \ref{app:SignalBackgroundComparison} for more detail.

\subsection{Imposing cuts}
\label{sec:ComparisonCuts}

There is no detector that is able to detect and reconstruct every muon. For instance, in the ATLAS detector at CERN the muon trigger chambers only cover the pseudorapidity region $|\eta| < 2.4$ \citep[chapter 6]{ATLAS_DesignReport}. In addition, muons are efficiently reconstructed only for high transverse momenta $p_T$. It is therefore sensible to impose cuts on the Monte-Carlo events. Starting from the event samples used in the last section and requiring
\begin{align}
	\label{eq:Cuts1}
	|\eta| &< 2.4, \\
	p_T &> 15 \text{ GeV}
\end{align}
should give a more realistic picture of MCPM discovery potential.

While this of course decreases statistics, there is one advantage. As discussed in section \ref{sec:SMBackground}, the spin-0 Higgs bosons are more likely to decay into this central region than the spin-1 $Z$ bosons and photons which make up the background. Hence these cuts reduce the background more drastically than the signal events.

\begin{figure}[!htb]
	\centering
	\subfloat[Raw data -- no cuts applied.]{\input{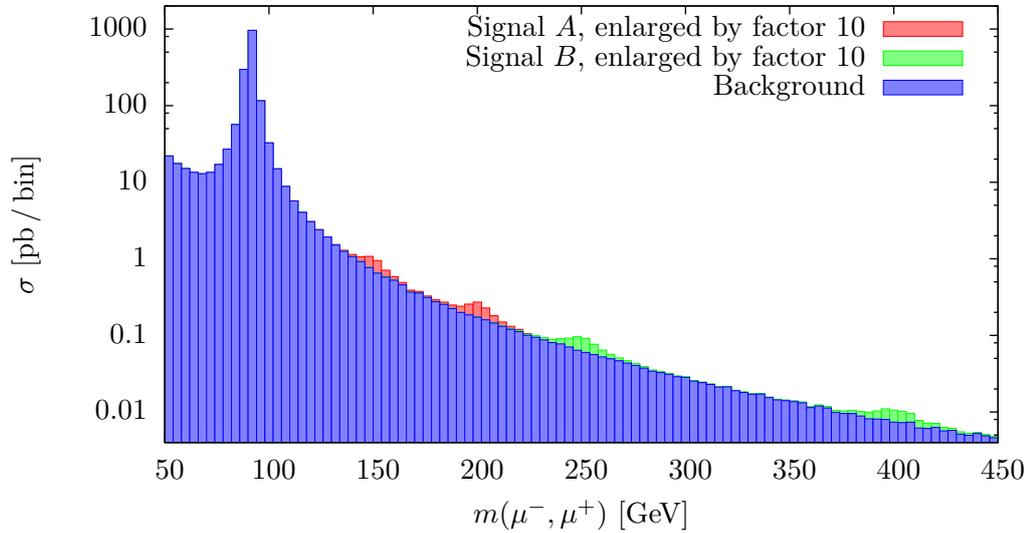} \label{fig:PlotComparisonBackgroundNoCuts14}}\\
	\subfloat[Imposing cuts on $\eta$ and $p_T$.]{\input{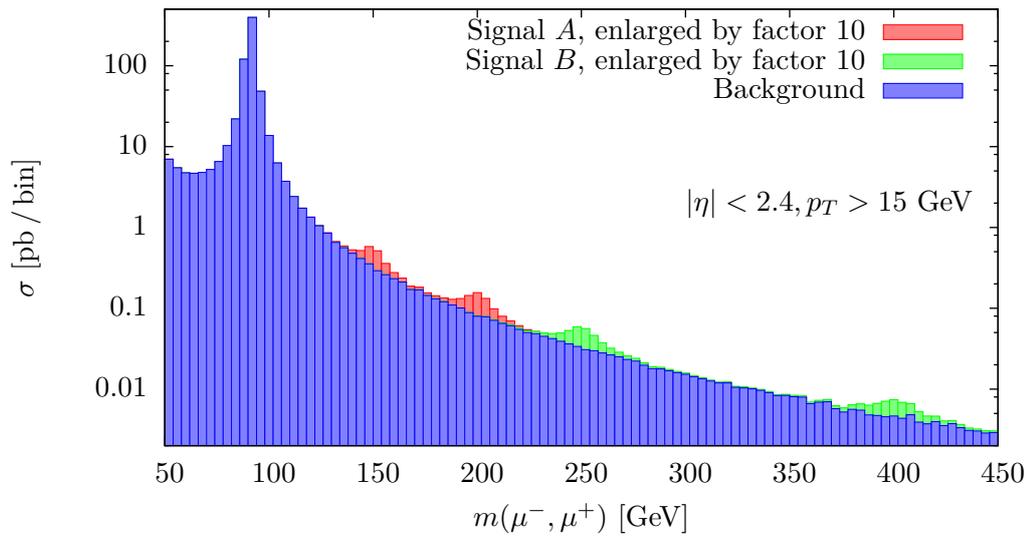} \label{fig:PlotComparisonBackgroundCuts14}}
	\caption{MCPM events compared to the SM background in the $\mu^- \mu^+$ channel as a function of the invariant mass of the muon pair. $\sqrt{s} = 14$ TeV. While the $Z$ resonance at 91.2 GeV is clearly visible, the Higgs resonances would be hard to notice without the magnification factor of 10.}
	\label{fig:PlotComparisonBackground14}
\end{figure}

Fig. \ref{fig:PlotComparisonBackgroundCuts14} shows the resulting invariant mass distribution. As expected, the MCPM signature is more clearly visible. For Higgs masses around 150 to 200 GeV and looking only at the resonance regions where the invariant mass of the muons is in the range of $m_H - \Gamma_H$ to $m_H + \Gamma_H$,
\begin{equation}
	\frac {\sigma_{\text{signal}}} {\sigma_{\text{background}}} \,\Big{|}_{\text{resonance}} \approx
	\begin{cases}
		5 \cdot 10^{-2} \quad \text{for } \sqrt{s} = 14 \text{ TeV} \\
		3 \cdot 10^{-2} \quad \text{for } \sqrt{s} = 7 \text{ TeV.} \\
	\end{cases}
	\label{eq:CrossSectionRatioCuts}
\end{equation}
Again, this value decreases for heavier Higgs bosons. More data can be found in appendix \ref{app:SignalBackgroundComparison}.

Now one can calculate how much statistics is needed so that the MCPM resonance peaks are \emph{locally} significant compared to the background, i.\,e.\ larger than $n \sigma$, where $\sigma$ is the statistical uncertainty of the background. Conventionally $n = 5$ is considered a discovery, while $n=2$ is often used for exclusion limits\footnote{These values are valid only locally, i.\,e.\ in the invariant mass range corresponding to the resonance peaks, and they cannot be simply translated into exclusion limits based on confidence levels. More sophisticated statistical tools are needed for this.}. I calculated the integrated luminosities $\int L \text d t$ needed for local $2 \sigma$ and $5 \sigma$ significance. The results are shown in table \ref{tbl:LuminositiesNeeded}.

\begin{table}[!htb]
	\centering
	\begin{tabular*}{\textwidth}{@{\extracolsep{\fill}} rrrrr}
	\toprule[0.12em]
	$m_H$ [GeV] & \multicolumn{4}{c}{$\int L \text d t$ [$\text{fb}^{-1}$]}\\
	\cmidrule{2-5}
	& \multicolumn{2}{c}{$2\sigma$ significance} & \multicolumn{2}{c}{$5\sigma$ significance}\\
	& $\sqrt{s} = 7$ TeV & $\sqrt{s} = 14$ TeV & $\sqrt{s} = 7$ TeV & $\sqrt{s} = 14$ TeV \\
	\midrule
	150 & 6.7 & 1.2 & 41.9 & 7.7 \\
	200 & 22.0 & 3.1 & 137.7 & 19.1 \\
	250 & 68.5 & 7.9 & 427.9 & 49.4 \\
	400 & 866.4 & 85.9 & 5415.0 & 536.8 \\
	\bottomrule[0.12em]
 	\end{tabular*}
	\caption{Integrated luminosities that are needed for local significances of the Higgs boson resonances.}
	\label{tbl:LuminositiesNeeded}
\end{table}

I have to stress though that these luminosities are very rough estimates. Only local significance in one channel is described, the combination of different peaks and different channels will improve the situation a bit. On the other hand, these numbers do not include detector efficiencies and resolution, only background and signal tree-level processes are taken into account. The experimental reality is much more complicated.

Keeping this caveat in mind, these numbers may nevertheless hint at the order of magnitude of statistics needed for MCPM signatures to become visible. As of October 31, 2011, the ATLAS detector at CERN collected a total of 5.25 $\text{fb}^{-1}$ of data at $\sqrt{s} = 7$ TeV \cite{ATLAS_Luminosity}. It is expected that the LHC will collect significantly more data in 2012 and will reach its design energy of $\sqrt{s} = 14$ TeV a few years later. Looking at the critical luminosities in table \ref{tbl:LuminositiesNeeded}, the next years will be an exciting time for the MCPM. Although a formal $5\sigma$ discovery seems very unlikely to happen soon, there might be either hints at the MCPM or exclusion limits on its Higgs boson mass space.

\section{A brief look at the $Z^\prime$ search}
\label{sec:ZPrimeSearch}

The MCPM is not the only extension of the SM that predicts $\mu^- \mu^+$ events, and due to reconstruction efficiency and low background this channel is used in various searches for signs of new physics, for instance $Z'$ boson resonances (for an introduction to $Z^\prime$ physics, see \citep{Rizzo_ZPrime}). In \citep{ATLAS_ZPrimeSearch}, 1.21 $\text{fb}^{-1}$ of data taken with the ATLAS detector at $\sqrt{s} = 7$ TeV were analysed for excesses over the SM predictions. In contrast to the estimates in the last section, the $Z'$ search includes a proper analysis of hadronisation, QCD background faking muons and detector properties.

In \citep{ATLAS_ZPrimeSearch}, the expected and measured number of events passing certain cuts were compared for different invariant mass bins. The most relevant cut is on the transverse momentum $p_T$: Each muon has to satisfy
\begin{equation}
	p_T > 25 \text{ GeV}.
\end{equation}
Due to detector geometry,
\begin{equation}
	|\eta| < 2.4
\end{equation}
is required. Other cuts limit the position of the primary vertex and the impact parameter.

Using MadGraph, the cross-sections for the usual $\mu^- \mu^+$ signatures involving $h'$ and $h''$ bosons were simulated. The parameters and cuts were chosen such as to closely resemble the parameters of the $Z'$ search\footnote{Some cuts could not be replicated in MadGraph, for instance restrictions on the primary vertex position.}. For the Higgs boson masses, two different sets of parameters $C$ and $D$ were chosen, see table \ref{tbl:ZPrimeComparisonHiggsMasses}. Notably, parameter set $C$ represents a best-case scenario with a large Higgs boson resonance peak at 120 GeV. The technical parameters were chosen as in section \ref{sec:ComparisonSignalBackground}.

\begin{table}[!htb]
	\centering
	\begin{tabular*}{\textwidth}{@{\extracolsep{\fill}} rrrrr}
	\toprule[0.12em]
	Parameter set & $m_{\rho'}$ [GeV] & $m_{h'}$ [GeV] & $m_{h''}$ [GeV] & $m_{H^\pm}$ [GeV]\\
	\midrule
	C &	125	& 120	& 120 & 125	\\
	D	& 125	& 250	& 150 & 150 \\
	\bottomrule[0.12em]
 	\end{tabular*}
	\caption{Parameter sets used for the comparison of $Z'$ search results and MCPM predictions.}
	\label{tbl:ZPrimeComparisonHiggsMasses}
\end{table}

Table \ref{tbl:ZPrimeResults} shows the results of the $Z'$ search and the MCPM predictions for the two settings. Even for the best-case scenario $C$ the MCPM predicts less of an excess than the statistical and systematical uncertainties of data and background analysis.

\begin{table}[!htb]
	\centering
	\begin{tabular}{l r@{$\,\pm\,$}r r@{$\,\pm\,$}r}
	\toprule[0.12em]
	$\mu^- \mu^+$ invariant mass bin [GeV] & \multicolumn{2}{c}{110--200} & \multicolumn{2}{c}{200--400} \\
	\midrule
	ATLAS data according to \citep{ATLAS_ZPrimeSearch} & 5406&74 & 557&24 \\
	SM prediction according to \citep{ATLAS_ZPrimeSearch} & 5434&150 & 571&23 \\
	\midrule
	MCPM signal (parameter set $C$) & 42&6 & 0&0 \\
	MCPM signal (parameter set $D$) & 19&4 & 3&2 \\
	\bottomrule[0.12em]
 	\end{tabular}
	\caption{Number of events measured at ATLAS compared to SM predictions and MCPM signatures. Here the invariant mass bins are relatively large, but even with smaller bins the MCPM signature is insignificant compared to the background uncertainties.}
	\label{tbl:ZPrimeResults}
\end{table}

A thorough analysis including detector simulation will further decrease the MCPM prediction. Thus it can be concluded that the data presented in \citep{ATLAS_ZPrimeSearch} is probably not sufficient to discover or exclude the MCPM for Higgs boson masses above 100 GeV. Of course, the combination with other channels and the analysis of the full data taken in 2011 might yield a different picture.

\chapter{Summary}
\label{ch:conclusions}

In this Bachelor thesis I implemented a two-Higgs-doublet model with maximal CP symmetry, the MCPM, into the Monte-Carlo event generation package MadGraph.

The MCPM is an extension of the Standard Model that is based on the requirement of invariance under generalised CP transformations. It features a unique fermion structure, in which only the third-generation fermions are massive. Of course this does not describe nature precisely, but the model gives a good approximation of what has been observed. The theory also predicts five Higgs bosons, of which four couple only to the second-generation fermions. At colliders such as the LHC, these Higgs bosons will be produced mainly via quark-antiquark fusion. Their $\mu^- \mu^+$ decay modes seem most promising for a discovery. For a proper search for MCPM signatures, a Monte-Carlo simulation is needed.

In the course of this thesis, the MCPM was implemented into MadGraph, allowing the calculation of cross-sections and the random generation of events for arbitrary tree-level processes. The output is in the LHEF format and can be used for further analysis, for instance using PYTHIA and GEANT. Therefore this MadGraph implementation can be used as the starting point for a full and thorough Monte-Carlo simulation of the MCPM. 

The implementation was validated successfully with different methods. It was then used to compare the MCPM signatures to the SM background at LHC energies. This analysis was only done for the $\mu^- \mu^+$ channel and restricted to hard processes, so the results are only a rough approximation. It was found that even after applying appropriate cuts, the Higgs boson resonance peaks are small compared to the background. For a Higgs boson of 150 GeV (200 GeV), 6.7 $\text{fb}^{-1}$ (22 $\text{fb}^{-1}$) of data at a center-of-mass energy $\sqrt{s} = 7$ TeV have to be collected so that a local $2\sigma$ significance becomes possible, not taking detector inefficiencies into account. It seems that the exclusion of a part of the Higgs mass parameter range or the discovery of small local excesses hinting at the MCPM might be within reach in the next years. Finally the MCPM predictions were compared to 2011 results of the $Z'$ search at the ATLAS detector, again hinting that more data must be collected in order to discover or exclude the MCPM for Higgs masses of more than about 100 GeV.

\cleardoublepage

\appendix

\chapter{Further details of the MCPM}
\label{app:MCPMSmallprint}

\section{The THDM Lagrangian}
\label{app:LagrangianBeforeSB}

The Lagrangian of a general THDM is given in \eqref{eq:THDMLagrangian} with the separate sectors given in \eqref{eq:THDMGaugeFermions}, \eqref{eq:THDMScalar} and \eqref{eq:THDMYukawa}. The MCPM is more specific, its Higgs potential and Yukawa sectors can be found in \eqref{eq:MCPMPotential} and \eqref{eq:MCPMYukawa}. In these terms, the following quantities appear:
\begin{itemize}
	\item $u^m$, $d^m$, $e^m$ and $\nu^m$ are fermionic fields representing up-type quarks, down-type quarks, charged leptons and lepton neutrinos. $m$ is a generation index, three generations have been observed in nature.
	\item With the projection operators
	\begin{equation}
		P_{L/R} = \frac {1 \mp \gamma_5} 2
	\end{equation}
	the left-handed and right-handed chiral projections of these fields can be defined as $\psi_{L/R} = P_{L/R} \psi$.
	\item The left-handed particles are organised in SU(2) doublets,
	\begin{equation}
		q_{L}^m = {\zvek {u^m} {d^m}}_L \text{ and } l_{L}^m = {\zvek {\nu^m} {e^m}}_L.
	\end{equation}
	\item There are two complex scalar doublets
	\begin{equation}
		\varphi_i(x) = \zvek {\varphi^+_i (x)} {\varphi^0_i (x)} \quad (i = 1,2).
	\end{equation}
	\item $G^i_\mu$, $W^i_\mu$ and $B_\mu$ are the gauge fields of the $SU(3)_C \times SU(2)_L \times U(1)_Y$ gauge group.
	\item $g_s$, $g$ and $g'$ are the corresponding coupling constants.
	\item Their field strength tensors are given by
	\begin{align}
		G_{\mu\nu}^i &= \partial_\mu G^i_\nu - \partial_\nu G^i_\mu - g_s f^{ijk} G^j_\mu G^k_\nu \\
		W_{\mu\nu}^i &= \partial_\mu W^i_\nu - \partial_\nu W^i_\mu - g \varepsilon^{ijk} W^j_\mu W^k_\nu \\
		B_{\mu\nu} &= \partial_\mu B_\nu - \partial_\nu B_\mu.
	\end{align}
	\item The covariant derivatives are
	\begin{align}
		D_\mu q_{L}^{m \alpha} &= \left( \partial_\mu  + i g_s L^{i} G^{i\alpha \beta}_\mu + i g T^i W^i_\mu \delta^{\alpha \beta} + i \frac {g'} 6 B_\mu  \delta^{\alpha \beta} \right) q_{L}^{m \beta} \\
		D_\mu l_{L}^m &= \left( \partial_\mu  + i g T^i W^i_\mu - i \frac {g'} 2 B_\mu \right) l_{L}^m \\
		D_\mu u_{R}^{m \alpha} &= \left( \partial_\mu  + i g_s L^{i} G^{i\alpha \beta}_\mu + i \frac {2g'} 3 B_\mu \delta^{\alpha \beta} \right) u_{R}^{m \beta} \\
		D_\mu d_{R}^{m \alpha} &= \left( \partial_\mu  + i g_s L^{i} G^{i\alpha \beta}_\mu - i \frac {g'} 3 B_\mu \delta^{\alpha \beta} \right) d_{R}^{m \beta} \\
		D_\mu e_{R}^m &= \left( \partial_\mu  - i g' B_\mu \right) e_{R}^m \\
		D_\mu \varphi_i &= \left( \partial_\mu  + i g T^j W^j_\mu + i \frac {g'} 2 B_\mu \right) \varphi_i.
		\label{eq:THDMCovDer}
	\end{align}
	\item $L^i$ and $T^i$ are $SU(3)$ and $SU(2)$ generators in the fundamental representation.
	\item $f^{ijk}$ and $\varepsilon^{ijk}$ are the structure constants of these representations.
	\item There are Higgs potential and Yukawa coefficients. In the general THDM these are called $a_{ij}$, $b_{klmn}$, $C_u^{imn}$, $C_d^{imn}$ and $C_l^{imn}$. In the MCPM with three generations the coefficients are $\mu_1$, $\mu_2$, $\mu_3$, $\xi_0$, $\eta_{00}$, $c_u$, $c_d$ and $c_l$.
	\item Colour and spinor indices are suppressed (with the exception of the quark colour indices $\alpha$ in the covariant derivative).
	\item Feynman slash notation is used:
	\begin{equation}
		\slashed{D} := \gamma^\mu D_\mu.
	\end{equation}
\end{itemize}

\section{Particle spectrum}
\label{app:MCPMParticles}

All the particles of the MCPM and their most important properties and couplings can be found in table \ref{tbl:MCPMFermions}, table \ref{tbl:MCPMGaugeBosons} and table \ref{tbl:MCPMHiggsBosons}.

\begin{table}[htb]
	\begin{tabular*}{\textwidth}{@{\extracolsep{\fill}} llrrll}
	\toprule[0.12em]
	Gen. & Fermion & Mass & Charge & \multicolumn{2}{c}{Couplings} \\
	\cmidrule{5-6}
	& & & & Gauge & Higgs \\
	\midrule
	\multirow{4}{*}{1} & $u$ & 0 & $\frac 2 3$ & $G$, $W^\pm$, $Z$, $\gamma$ & -\\
	& $d$ & 0 & $- \frac 1 3$ &  $G$, $W^\pm$, $Z$, $\gamma$ & -\\
	& $e$ & 0 & $-1$ &  $W^\pm$, $Z$, $\gamma$ & -\\
	& $\nu_e$ & 0 & 0 &  $W^\pm$, $Z$ & - \\
	\midrule
	\multirow{4}{*}{2} & $c$ & 0 & $\frac 2 3$ & $G$, $W^\pm$, $Z$, $\gamma$ & $h'$, $h''$, $H^\pm$\\
	& $s$ & 0 & $- \frac 1 3$ &  $G$, $W^\pm$, $Z$, $\gamma$ & $h'$, $h''$, $H^\pm$\\
	& $\mu$ & 0 & $-1$ &  $W^\pm$, $Z$, $\gamma$ & $h'$, $h''$, $H^\pm$\\
	& $\nu_\mu$ & 0 & 0 &  $W^\pm$, $Z$ & $h'$, $h''$, $H^\pm$\\
	\midrule
	\multirow{4}{*}{3} & $t$ & $m_t$ & $\frac 2 3$ & $G$, $W^\pm$, $Z$, $\gamma$ & $\rho'$\\
	& $b$ & $m_b$ & $- \frac 1 3$ &  $G$, $W^\pm$, $Z$, $\gamma$ & $\rho'$\\
	& $\tau$ & $m_\tau$ & $-1$ &  $W^\pm$, $Z$, $\gamma$ & $\rho'$\\
	& $\nu_\tau$ & 0 & 0 &  $W^\pm$, $Z$ & $\rho'$ \\
	\bottomrule[0.12em]
	\end{tabular*}
	\caption{Fermions in the MCPM, their key properties and couplings. Unlike in the SM, the first two generation of fermions are massless. The third-generation fermion masses are given in \eqref{eq:MCPMFermionMasses1} to \eqref{eq:MCPMFermionMasses2}.}
	\label{tbl:MCPMFermions}
\end{table}

\begin{table}[htb]
	\begin{tabular*}{\textwidth}{@{\extracolsep{\fill}} lrrlll}
	\toprule[0.12em]
	Boson & Mass & Charge & \multicolumn{3}{c}{Couplings} \\
	\cmidrule{4-6}
	& & & Fermions & Gauge & Higgs \\
	\midrule
	$G$ & 0 & 0 & Quarks & $G$ & - \\
	$W^\pm$ & $m_W$ & $\pm 1$ & All fermions & $W^\pm$, $Z$, $\gamma$ & $\rho'$, $h'$, $h''$, $H^\pm$\\
	$Z$ & $m_Z$ & 0 & All fermions & $W^\pm$, $Z$, $\gamma$ & $\rho'$, $h'$, $h''$, $H^\pm$ \\
	$\gamma$ & 0 & 0 & Quarks, charged leptons & $W^\pm$, $Z$, $\gamma$ & $h'$, $h''$, $H^\pm$ \\
	\bottomrule[0.12em]
	\end{tabular*}
	\caption{Gauge bosons in the MCPM. The gauge bosons and their couplings to the fermions are exactly the same as in the SM.}
	\label{tbl:MCPMGaugeBosons}
\end{table}

\begin{table}[htb]
	\begin{tabular*}{\textwidth}{@{\extracolsep{\fill}} lrrlll}
	\toprule[0.12em]
	Boson & Mass & Charge & \multicolumn{3}{c}{Couplings} \\
	\cmidrule{4-6}
	& & & Fermions & Gauge & Higgs \\
	\midrule
	$\rho'$ & $m_{\rho'}$ & 0 & $t$, $b$, $\tau$ & $W^\pm$, $Z$ & all \\
	$h'$ & $m_{h'}$ & 0 & $c$, $s$, $\mu$ & $W^\pm$, $Z$, $\gamma$ & all \\
	$h''$ & $m_{h''}$ & 0 & $c$, $s$, $\mu$ & $W^\pm$, $Z$, $\gamma$ & all \\
	$H^\pm$ & $m_{H^\pm}$ & $\pm 1$ & $c$, $s$, $\mu$, $\nu_\mu$ & $W^\pm$, $Z$, $\gamma$ & all \\
	\bottomrule[0.12em]
	\end{tabular*}
	\caption{Higgs bosons in the MCPM. While the $\rho'$ is similar to the SM Higgs boson, the other Higgs particles have very distinctive properties. The Higgs boson masses are given in \eqref{eq:MCPMHiggsBosonMasses1} to \eqref{eq:MCPMHiggsBosonMasses2}. Note that some of the couplings are four-point vertices. For instance, the uncharged $h'$ and $h''$ couple to the $\gamma$ only via vertices involving $W^\pm$ and $H^\pm$ bosons as well.}
	\label{tbl:MCPMHiggsBosons}
\end{table}

\ \\

\section{Feynman rules}
\label{app:FeynmanRules}

In \citep[appendix A]{Maniatis_MCPMPhenomenology} the MCPM Lagrangian after symmetry breaking is presented and the Feynman rules for several vertices are derived. The most important vertices and the corresponding expressions are given in table \ref{tbl:FeynmanRules1} and table \ref{tbl:FeynmanRules2} below. 

\begin{table}[p]
		\centering
		\begin{tabular*}{.75 \textwidth}{@{\extracolsep{\fill}} cc}
			\toprule[0.12em]
			Vertex & Corresponding expression \\
			\midrule
			
			\quad & \quad \\
			\quad \quad	\begin{fmffile}{Feynman1}
				\begin{fmfgraph*}(60,40)
					\fmfleft{i}
					\fmfright{o2,o1}
					\fmflabel{$t$}{i}
					\fmflabel{$\rho'$}{o1}
					\fmflabel{$t$}{o2}
					\fmf{fermion}{i,v}
					\fmf{dashes}{v,o1}
					\fmf{fermion}{v,o2}
				\end{fmfgraph*}
			\end{fmffile} \quad \quad
			& $\displaystyle -i \frac {m_t} {v_0}$ \\
			\quad & \quad \\
			
			\quad & \quad \\
			\quad \quad	\begin{fmffile}{Feynman2}
				\begin{fmfgraph*}(60,40)
					\fmfleft{i}
					\fmfright{o2,o1}
					\fmflabel{$b$}{i}
					\fmflabel{$\rho'$}{o1}
					\fmflabel{$b$}{o2}
					\fmf{fermion}{i,v}
					\fmf{dashes}{v,o1}
					\fmf{fermion}{v,o2}
				\end{fmfgraph*}
			\end{fmffile} \quad \quad
			& $\displaystyle -i \frac {m_b} {v_0}$ \\
			\quad & \quad \\
			
			\quad & \quad \\
			\quad \quad	\begin{fmffile}{Feynman3}
				\begin{fmfgraph*}(60,40)
					\fmfleft{i}
					\fmfright{o2,o1}
					\fmflabel{$\tau$}{i}
					\fmflabel{$\rho'$}{o1}
					\fmflabel{$\tau$}{o2}
					\fmf{fermion}{i,v}
					\fmf{dashes}{v,o1}
					\fmf{fermion}{v,o2}
				\end{fmfgraph*}
			\end{fmffile} \quad \quad
			& $\displaystyle -i \frac {m_\tau} {v_0}$ \\
			\quad & \quad \\
			
			\quad & \quad \\
			\quad \quad	\begin{fmffile}{Feynman4}
				\begin{fmfgraph*}(60,40)
					\fmfleft{i}
					\fmfright{o2,o1}
					\fmflabel{$c$}{i}
					\fmflabel{$h'$}{o1}
					\fmflabel{$c$}{o2}
					\fmf{fermion}{i,v}
					\fmf{dashes}{v,o1}
					\fmf{fermion}{v,o2}
				\end{fmfgraph*}
			\end{fmffile} \quad \quad
			& $\displaystyle i \frac {m_t} {v_0}$ \\
			\quad & \quad \\
			
			\quad & \quad \\
			\quad \quad	\begin{fmffile}{Feynman5}
				\begin{fmfgraph*}(60,40)
					\fmfleft{i}
					\fmfright{o2,o1}
					\fmflabel{$s$}{i}
					\fmflabel{$h'$}{o1}
					\fmflabel{$s$}{o2}
					\fmf{fermion}{i,v}
					\fmf{dashes}{v,o1}
					\fmf{fermion}{v,o2}
				\end{fmfgraph*}
			\end{fmffile} \quad \quad
			& $\displaystyle i \frac {m_b} {v_0}$ \\
			\quad & \quad \\
			
			\quad & \quad \\
			\quad \quad	\begin{fmffile}{Feynman6}
				\begin{fmfgraph*}(60,40)
					\fmfleft{i}
					\fmfright{o2,o1}
					\fmflabel{$\mu$}{i}
					\fmflabel{$h'$}{o1}
					\fmflabel{$\mu$}{o2}
					\fmf{fermion}{i,v}
					\fmf{dashes}{v,o1}
					\fmf{fermion}{v,o2}
				\end{fmfgraph*}
			\end{fmffile} \quad \quad
			& $\displaystyle i \frac {m_\tau} {v_0}$ \\
			\quad & \quad \\
			
			\bottomrule[0.12em]
		\end{tabular*}
	\caption{Feynman rules for the most relevant vertices in the MCPM, part 1.}
	\label{tbl:FeynmanRules1}
\end{table}

\begin{table}[p]
		\centering
		\begin{tabular*}{.75 \textwidth}{@{\extracolsep{\fill}} cc}
			\toprule[0.12em]
			Vertex & Corresponding expression \\
			\midrule
			
			\quad & \quad \\
			\quad \quad	\begin{fmffile}{Feynman7}
				\begin{fmfgraph*}(60,40)
					\fmfleft{i}
					\fmfright{o2,o1}
					\fmflabel{$c$}{i}
					\fmflabel{$h''$}{o1}
					\fmflabel{$c$}{o2}
					\fmf{fermion}{i,v}
					\fmf{dashes}{v,o1}
					\fmf{fermion}{v,o2}
				\end{fmfgraph*}
			\end{fmffile} \quad \quad
			& $\displaystyle \frac {m_t} {v_0} \gamma_5$ \\
			\quad & \quad \\
			
			\quad & \quad \\
			\quad \quad	\begin{fmffile}{Feynman8}
				\begin{fmfgraph*}(60,40)
					\fmfleft{i}
					\fmfright{o2,o1}
					\fmflabel{$s$}{i}
					\fmflabel{$h''$}{o1}
					\fmflabel{$s$}{o2}
					\fmf{fermion}{i,v}
					\fmf{dashes}{v,o1}
					\fmf{fermion}{v,o2}
				\end{fmfgraph*}
			\end{fmffile} \quad \quad
			& $\displaystyle - \frac {m_b} {v_0} \gamma_5$ \\
			\quad & \quad \\
			
			\quad & \quad \\
			\quad \quad	\begin{fmffile}{Feynman9}
				\begin{fmfgraph*}(60,40)
					\fmfleft{i}
					\fmfright{o2,o1}
					\fmflabel{$\mu$}{i}
					\fmflabel{$h''$}{o1}
					\fmflabel{$\mu$}{o2}
					\fmf{fermion}{i,v}
					\fmf{dashes}{v,o1}
					\fmf{fermion}{v,o2}
				\end{fmfgraph*}
			\end{fmffile} \quad \quad
			& $\displaystyle - \frac {m_\tau} {v_0} \gamma_5$ \\
			\quad & \quad \\
						
			\quad & \quad \\
			\quad \quad	\begin{fmffile}{Feynman10}
				\begin{fmfgraph*}(60,40)
					\fmfleft{i}
					\fmfright{o2,o1}
					\fmflabel{$s$}{i}
					\fmflabel{$H^-$}{o1}
					\fmflabel{$c$}{o2}
					\fmf{fermion}{i,v}
					\fmf{dashes_arrow}{v,o1}
					\fmf{fermion}{v,o2}
				\end{fmfgraph*}
			\end{fmffile} \quad \quad
			& $\displaystyle - i \frac {1} {\sqrt{2} v_0} \left[ m_t (1 - \gamma_5) - m_b (1 + \gamma_5) \right]$ \\
			\quad & \quad \\
			
			\quad & \quad \\
			\quad \quad	\begin{fmffile}{Feynman11}
				\begin{fmfgraph*}(60,40)
					\fmfleft{i}
					\fmfright{o2,o1}
					\fmflabel{$\mu^-$}{i}
					\fmflabel{$H^-$}{o1}
					\fmflabel{$\nu_\mu$}{o2}
					\fmf{fermion}{i,v}
					\fmf{dashes_arrow}{v,o1}
					\fmf{fermion}{v,o2}
				\end{fmfgraph*}
			\end{fmffile} \quad \quad
			& $\displaystyle i \frac {m_\tau} {\sqrt{2} v_0} (1 + \gamma_5)$ \\
			\quad & \quad \\
						
			\bottomrule[0.12em]
		\end{tabular*}
	\caption{Feynman rules for the most relevant vertices in the MCPM, part 2. There are many more, including couplings of gauge bosons to Higgs bosons and Higgs boson self-couplings. The arrows on the $H^\pm$ lines denote the flow of negative charge.}
	\label{tbl:FeynmanRules2}
\end{table}

\section{Parameters and their relations}
\label{app:MCPMParameters}

As described in section \ref{sec:ParameterSpace}, the independent parameters used in the analysis are $\alpha$, $G_F$, $\alpha_s$, $m_Z$, $m_t$, $m_b$, $m_\tau$, $m_{\rho'}$, $m_{h'}$, $m_{h''}$ and $m_{H^\pm}$. All other quantities can be calculated from these parameters as follows \citep[appendix A]{Maniatis_MCPMPhenomenology}:
\begin{align}
	\xi_0 &= - \frac 1 2 m_{\rho'}^2, \\ \displaybreak[1]
	\eta_{00} &= \frac 1 {2v_0^2} \left(m_{H^\pm}^2 + m_{\rho'}^2 \right), \\ \displaybreak[1]
	\mu_1 &= \frac 1 {2v_0^2} \left(m_{h'}^2 - m_{H^\pm}^2 \right), \\ \displaybreak[1]
	\mu_2 &= \frac 1 {2v_0^2} \left(m_{h''}^2 - m_{H^\pm}^2 \right), \\ \displaybreak[1]
	\mu_3 &= - \frac 1 {2v_0^2} m_{H^\pm}^2, \\ \displaybreak[1]
	e &= \sqrt{4 \pi \alpha}, \\ \displaybreak[1]
	\sin^2 \theta_W &= \frac 1 2 \left[ 1 - \left( 1 - \frac {e^2} {\sqrt{2} G_F m_Z^2} \right)^{\frac 1 2} \right], \\ \displaybreak[1]
	m_W^2 &= \frac {m_Z^2} 2 \left[ 1 + \left( 1 - \frac {e^2} {\sqrt{2} G_F m_Z^2} \right)^{\frac 1 2} \right], \\ \displaybreak[1]
	v_0 &= 2^{- \frac 1 4} G_F^{- \frac 1 2}.
\end{align}

\section{Higgs boson decay modes}
\label{app:DecayCoefficients}

In \eqref{eq:HiggsDecayWidth} the width of Higgs decays into fermion-antifermion pairs was given, including coefficients that are specific to the different decay modes. These coefficients have been calculated by \citet[table 2]{Maniatis_MCPMPhenomenology} and are listed in table \ref{tbl:HiggsDecayCoefficients} below.
\begin{table}[!htb]
		\centering
		\begin{tabular*}{\textwidth}{@{\extracolsep{\fill}} llrrr}
			\toprule[0.12em]
			Higgs boson & Decay mode & $a$ & $b$ & $N_c$ \\
			\midrule
			$\rho'$ & $t \bar t$ & $m_t$ & 0 & 3 \\
			& $b \bar b$ & $m_b$ & 0 & 3\\
			& $\tau^- \tau^+$ & $m_\tau$ & 0 & 1 \\
			\midrule
			$h'$ & $c \bar c$ & $- m_t$ & 0 & 3 \\
			& $s \bar s$ & $- m_b$ & 0 & 3\\
			& $\mu^- \mu^+$ & $- m_\tau$ & 0 & 1 \\
			\midrule
			$h''$ & $c \bar c$ & 0 & $i m_t$ & 3 \\
			& $s \bar s$ & 0 & $- i m_b$ & 3\\
			& $\mu^- \mu^+$ & 0 & $- i m_\tau$ & 1 \\
			\midrule
			$H^+$ & $c \bar s$ & $\frac {m_t - m_b} {\sqrt 2}$ & $-\frac {m_t + m_b} {\sqrt 2}$ & 3 \\
			& $\mu^+ \nu_\mu$ & $- \frac {m_\tau} {\sqrt 2}$ & $- \frac {m_\tau} {\sqrt 2}$ & 1 \\
			\midrule
			$H^-$ & $s \bar c$ & $\frac {m_t - m_b} {\sqrt 2}$ & $\frac {m_t + m_b} {\sqrt 2}$ & 3 \\
			& $\mu^- \bar{\nu}_\mu$ & $- \frac {m_\tau} {\sqrt 2}$ & $\frac {m_\tau} {\sqrt 2}$ & 1 \\	
			\bottomrule[0.12em]
		\end{tabular*}
	\caption{Higgs boson fermionic decay modes and their coefficients. The decay widths for these modes are given in \eqref{eq:HiggsDecayWidth}.}
	\label{tbl:HiggsDecayCoefficients}
\end{table}

\chapter{Technical aspects of the MadGraph implementation}
\label{app:implementation}

\section{Instructions on usage}
\label{app:ImplementationUsage}

The MCPM model in MadGraph is used in the same way as every MadGraph model, see \citep{MadGraph5} for an overview. First it has to be loaded. After starting the MadGraph binary in a shell this can be done with the command
\begin{verbatim}
	import model MCPM -modelname
\end{verbatim}
where the option \verb|-modelname| is needed to ensure the correct particle names. Then processes are defined and the event generation started:
\begin{verbatim}
	generate p p > h1 > m- m+
	add process p p > h2 > m- m+
	output -f
	launch
\end{verbatim}
The \verb|generate| and (optionally) \verb|add process| commands are used to specify the processes MadGraph has to evaluate, while \verb|output| and \verb|launch| launch the event generation. The particle names used in the implementation are given in the following section. Before events are produced, MadGraph asks the user whether the standard parameters are to be used or if he wants to modify them.

During the simulation, an in-browser status page keeps the user informed. In the end, MadGraph gives out the contributing Feynman diagrams, the cross-sections or decay widths, the corresponding uncertainty and a compressed file containing the generated events in LHEF \citep{LHEF} format. This can be used for further analysis, for instance with MadAnalysis or PYTHIA \citep{PYTHIA}.

\section{List of particles and parameters}
\label{app:ImplementationParticlesParameters}

A list of MCPM particles and parameters with their MadGraph names are presented in table \ref{tbl:ImplementationParticles} and \ref{tbl:ImplementationParameters} respectively.

\begin{table}[!htb]
	\centering
		\begin{tabular*}{\textwidth}{@{\extracolsep{\fill}} ll}
			\toprule[0.12em]
			Particles & Names in MadGraph MCPM implementation \\
			\midrule
			Protons & \verb|p| \\
			\midrule
			Quarks & \verb|u d c s t b| \\
			Antiquarks & \verb|u~ d~ c~ s~ t~ b~| \\
			Leptons & \verb|e- m- tt- ve vm vt| \\
			Antileptons & \verb|e+ m+ tt+ ve~ vm~ vt~| \\
			\midrule
			Gauge bosons & \verb|a z w+ w- g| \\
			\midrule
			$\rho'$, $h'$, $h''$ & \verb|rho h1 h2| \\
			$H^\pm$ & \verb|h+ h-| \\
			\bottomrule[0.12em]
		\end{tabular*}
	\caption{The MCPM particles and their names in the MadGraph implementation.}
	\label{tbl:ImplementationParticles}
\end{table}

\begin{table}[!htb]
	\centering
		\begin{tabular*}{\textwidth}{@{\extracolsep{\fill}} lllrlr}
			\toprule[0.12em]
			Section & Parameter & MadGraph & Default & Unit & Strict MCPM\\
			\midrule
			Gauge & $\frac 1 \alpha$ & \verb|aEWM1| & 127.916 & &\\
			& $G_F$ & \verb|Gf| & 0.000011664 & $\text{GeV}^{-2}$ &\\
			& $\alpha_s$ & \verb|aS| & 0.1184 & &\\
			& $m_Z$ & \verb|MZ| & 91.1876 & GeV & \\
			\midrule
			CKM matrix & $\lambda_{WS}$ & \verb|lamWS| & 0.2253 & & 0\\
			& $A_{WS}$ & \verb|AWS| & 0.808 & &\\
			& $\rho_{WS}$ & \verb|rhoWS| & 0.132 & &\\
			& $\eta_{WS}$ & \verb|etaWS| & 0.341 & &\\
			\midrule
			Fermion masses & $m_c$ & \verb|MC| & 1.42 & GeV & 0 \\
			& $m_t$ & \verb|MT| & 172.9 & GeV & \\
			& $m_b$ & \verb|MB| & 4.67 & GeV & \\
			& $m_e$ & \verb|Me| & 0.00051100 & GeV & 0 \\
			& $m_\mu$ & \verb|MM| & 0.10566 & GeV & 0 \\
			& $m_\tau$ & \verb|MTA| & 1.7768 & GeV & \\
			\midrule
			Higgs masses & $m_{\rho'}$ & \verb|mrho| & 125  & GeV & \\
			& $m_{h'}$ & \verb|mh1| & 200 & GeV & \\
			& $m_{h''}$ & \verb|mh2| & 150 & GeV & \\
			& $m_{H^\pm}$ & \verb|mhc| & 150 & GeV & \\			
			\bottomrule[0.12em]
		\end{tabular*}
	\caption{The MCPM parameters and their names in the MadGraph implementation. Note that some parameters are zero in the MCPM (see last column), however they can be set to other values in this model. This allows the correct simulation of SM background processes (see section \ref{sec:MCImplementation}). The CKM matrix is given in the Wolfenstein parametrisation \citep[section 7.8.1]{Berger}.}
	\label{tbl:ImplementationParameters}
\end{table}

The parameters may be a bit confusing, because the $c$, $e$, $\mu$ can be set to be massive, while the $u$, $d$ and $s$ are always massless. This does not have anything to do with the MCPM. As explained in section \ref{sec:MCImplementation}, some of the consequences of strict symmetry under generalised CP transformations have been dropped in this implementation, so it is generally possible to set masses for the first- and second-generation fermions. However, MadGraph treats the very light fermions as massless (even in the SM, where the theory certainly predicts something else). This does not change the physics at collider experiments in the TeV range, but it saves some computation time. Therefore the list of parameters includes the masses of \emph{some} fermions of the first two generations.

MadGraph also allows to set decay widths for all particles, which are named similarly to the mass parameters. This is important for further analysis with programs such as PYTHIA, but these parameters do not influence the generation of events with MadGraph at all.

\chapter{Results of the validation}
\label{app:validation}

\section{Production cross-sections and decay widths}
\label{app:SystematicCheck}

The first check of the Monte-Carlo simulation consisted of the calculation of cross-sections and decay widths, see section \ref{sec:SystematicCheck}. The results are listed in table \ref{tbl:SystematicCheckResultsProduction} for the Higgs boson production and in table \ref{tbl:SystematicCheckResultsDecays} for the decay.

\begin{table}[!htb]
	\centering
		\begin{tabular}{ lrr     r@{.}l @{\,$\pm$\,} r@{.}l    r@{.}l }
			\toprule[0.12em]
			Process & Parameter set & $m_H$ [GeV] & \multicolumn{4}{c}{$\sigma_{\text{MadGraph}}$ [pb]} & \multicolumn{2}{c}{$\sigma_{\text{theo}}$ [pb]} \\
			\midrule
			$p + p \rightarrow h' + X$ & 1 & 100 & 13188&0 & 8&0	& 13195&0 \\
			& 2 & 400 & 75&382 & 0&048 & 75&396 \\
			& 3 & 200 & 1158&7 & 0&7 & 1159&0 \\
			& 4 & 300 & 244&44 & 0&15 & 244&50 \\
			\midrule
			$p + p \rightarrow h'' + X$ & 1 & 100 & 13199&0 & 8&0	& 13195&0 \\
			& 2 & 200 & 1159&0 & 0&7 & 1159&0 \\
			& 3 & 150 & 3284&7 & 2&0 & 3284&5 \\
			& 4 & 250 & 499&06 & 0&31 & 499&10 \\
			\midrule
			$p + p \rightarrow H^- + X$ & 1 & 150 & 4454&7 & 2&6	& 4454&5 \\
			& 2 & 250 & 722&89 & 0&44 & 722&87 \\
			& 3 & 200 & 1644&4 & 1&0 & 1644&4 \\
			& 4 & 300 & 360&96 & 0&22 & 360&95 \\
			\bottomrule[0.12em]
		\end{tabular}
	\caption{The results of the systematic check of production cross-sections. Parameter sets and other details can be found in section \ref{sec:SystematicCheckMethodology}. The deviations are very small, to say the least.}
	\label{tbl:SystematicCheckResultsProduction}
\end{table}

\begin{table}[!htb]
	\centering
	\subfloat[Hadronic decays.]{
		\begin{tabular}{ p{3cm}rr     r@{.}l @{$\,\pm\,$} r@{.}l    r@{.}l }
			\toprule[0.12em]
			Process & Parameter set & $m_H$ [GeV] & \multicolumn{4}{c}{$\Gamma_{\text{MadGraph}}$ [GeV]} & \multicolumn{2}{c}{$\Gamma_{\text{theo}}$ [GeV]} \\
			\midrule
			$h' \rightarrow c + \bar c$ & 1 & 100 & 5&8789 & 0&0072	& 5&8789 \\
			& 2 & 400 & 23&542 & 0&029 & 23&542 \\
			& 3 & 200 & 11&769 & 0&014 & 11&769 \\
			& 4 & 300 & 17&656 & 0&022 & 17&656 \\
			\midrule
			$h'' \rightarrow c + \bar c$ & 1 & 100 & 5&8837 & 0&0072	& 5&8837 \\
			& 2 & 200 & 11&771 & 0&014 & 11&771 \\
			& 3 & 150 & 8&8275 & 0&0107 & 8&8275 \\
			& 4 & 250 & 14&714 & 0&018 & 14&714 \\
			\midrule
			$H^- \rightarrow s + \bar c$ & 1 & 150 & 8&8339 & 0&0108	& 8&8339 \\
			& 2 & 250 & 14&725 & 0&018 & 14&725 \\
			& 3 & 200 & 11&779 & 0&014 & 11&779 \\
			& 4 & 300 & 17&670 & 0&022 & 17&670 \\
			\bottomrule[0.12em]
		\end{tabular}
		\label{tbl:SystematicCheckResultsHadronicDecays}}\\
	\subfloat[Leptonic Decays.]{
		\begin{tabular}{ p{3cm}rr     r@{.}l @{$\pm$} r@{.}l    r@{.}l }
			\toprule[0.12em]
			Process & Parameter set & $m_H$ [GeV] & \multicolumn{4}{c}{$\Gamma_{\text{MadGraph}}$ [keV]} & \multicolumn{2}{c}{$\Gamma_{\text{theo}}$ [keV]} \\
			\midrule
			$h' \rightarrow \mu^- + \mu^+$ & 1 & 100 & 207&20 \ & \ 0&25	& 207&20 \\
			& 2 & 400 & 828&82 & 1&01 & 828&82 \\
			& 3 & 200 & 414&41 & 0&51 & 414&41 \\
			& 4 & 300 & 621&61 & 0&76 & 621&61 \\
			\midrule
			$h'' \rightarrow \mu^- + \mu^+$ & 1 & 100 & 207&20 \ & \ 0&25	& 207&20 \\
			& 2 & 200 & 414&41 & 0&51 & 414&41 \\
			& 3 & 150 & 310&81 & 0&38 & 310&81 \\
			& 4 & 250 & 518&01 & 0&63 & 518&01 \\
			\midrule
			$H^- \rightarrow \mu^- + \bar{\nu}_\mu$ & 1 & 150 & 310&81 & 0&38 & 310&81 \\
			& 2 & 250 & 518&01 & 0&63 & 518&01 \\
			& 3 & 200 & 414&41 & 0&51 & 414&41 \\
			& 4 & 300 & 621&61 & 0&76 & 621&61 \\
			\bottomrule[0.12em]
		\end{tabular}
		\label{tbl:SystematicCheckResultsLeptonicDecays}}
	\caption{The results of the systematic check of total decay widths. Monte-Carlo results and theory agree perfectly.}
	\label{tbl:SystematicCheckResultsDecays}
\end{table}

\section{Full processes: Additional plots}
\label{app:FullProcesses}

\begin{figure}[htb]
	\centering
	\input{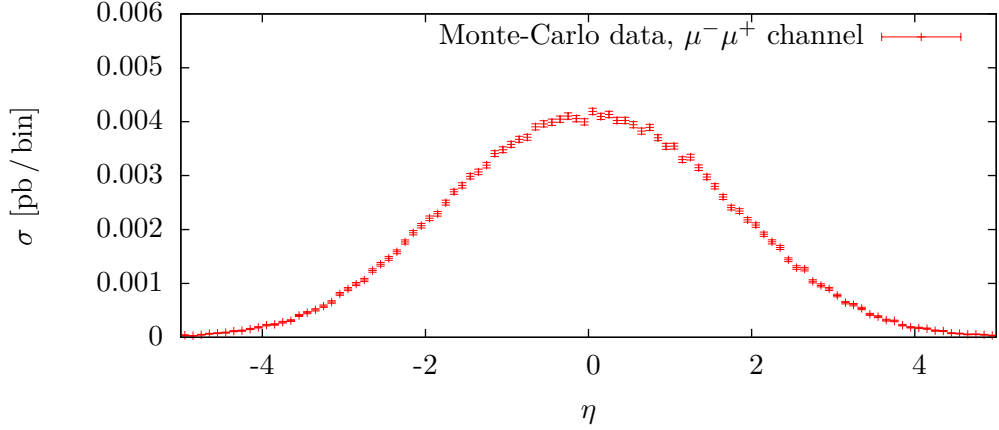}
	\caption{Dependency of the $\mu^- \mu^+$ cross-section on the pseudorapidity $\eta$ of the $\mu^-$ in the lab frame. $\sqrt{s} = 14$ TeV.}
	\label{fig:PlotEtaMuMu}
\end{figure}

\begin{figure}[htb]
	\centering
	\input{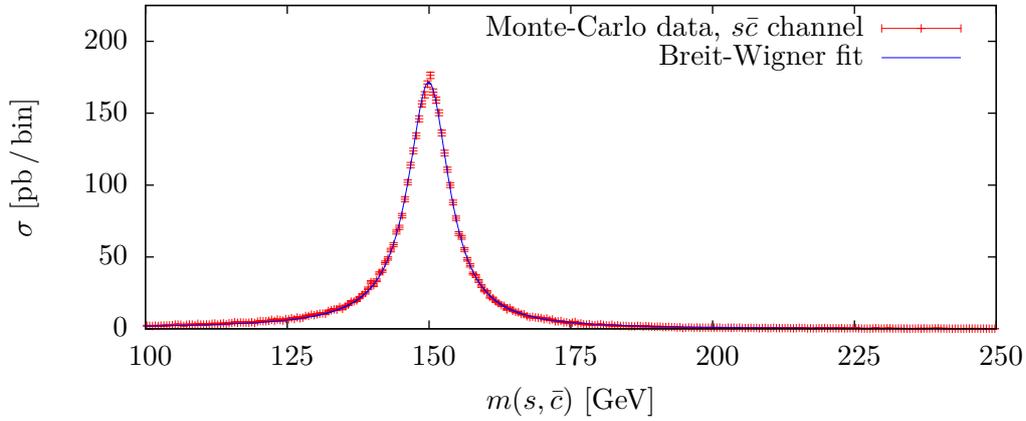}
	\caption{Resonance in the $s \bar c$ channel with a Breit-Wigner fit. $\sqrt{s} = 14$ TeV.}
	\label{fig:PlotResonanceShapeSC}
\end{figure}

Fig. \ref{fig:PlotEtaMuMu} shows the distribution of pseudorapidity $\eta$ in the lab frame for the $\mu^- \mu^+$ channel. In fig.\ \ref{fig:PlotResonanceShapeSC} the resonance peak for the $s \bar c$ mode can be seen. In either case, the processes and parameters described in section \ref{sec:FullProcessesMethodology} were used.

\chapter{Results of the analysis of MCPM signatures}
\label{app:analysis}

\section{Comparison of MCPM signal with background}
\label{app:SignalBackgroundComparison}

\begin{figure}[htb]
	\centering
	\subfloat[Raw data -- no cuts applied.]{\input{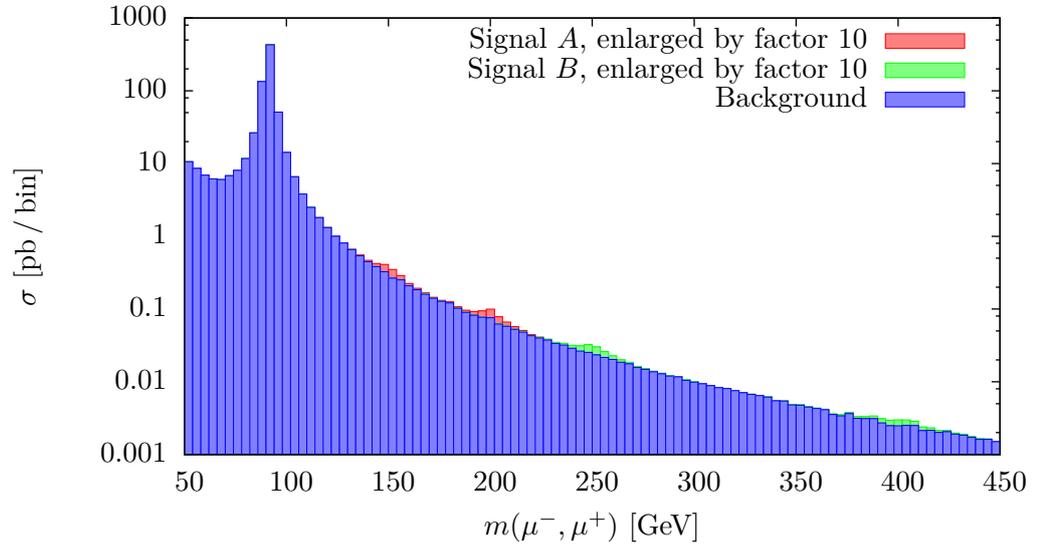} \label{fig:PlotComparisonBackgroundNoCuts7}}\\
	\subfloat[After applying cuts on $\eta$ and $p_T$.]{\input{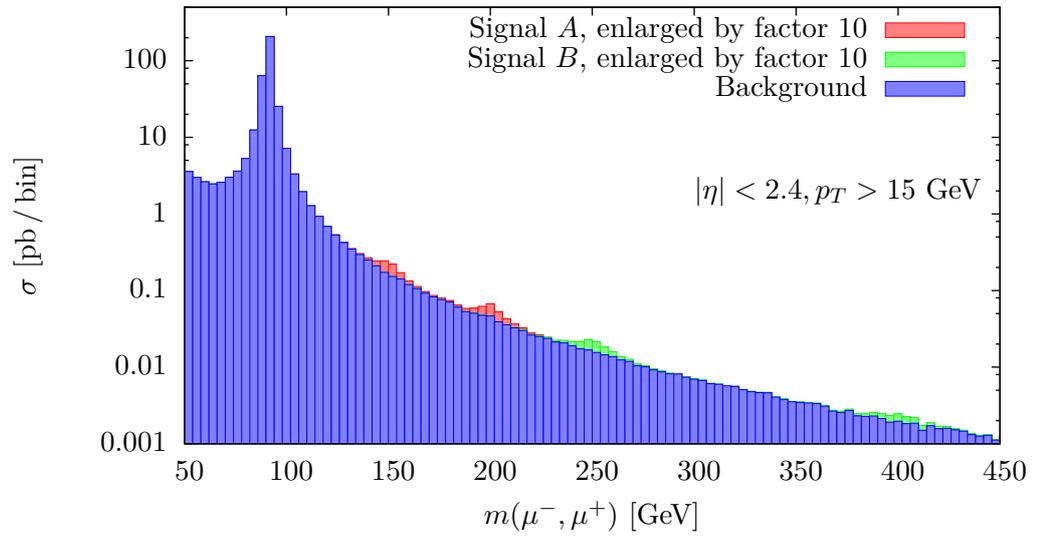} \label{fig:PlotComparisonBackgroundCuts7}}
	\caption{MCPM events compared to the SM background in the $\mu^- \mu^+$ channel as a function of the invariant mass of the muon pair. $\sqrt{s} = 7$ TeV.}
	\label{fig:PlotComparisonBackground7}
\end{figure}

In section \ref{sec:ComparisonSignalBackground}, the MCPM $\mu^- \mu^+$ signature was compared to the SM background. Fig. \ref{fig:PlotComparisonBackground7} shows the cross-sections of MCPM and SM as a function of the invariant mass of the muons for $\sqrt{s} = 7$ TeV, both with and without cuts. In table \ref{tbl:ComparisonResonancesBackground} the size of the Higgs resonance peaks is compared to the background for a variety of settings.

\begin{table}[!htb]
	\centering
	\subfloat[Raw data -- no cuts applied.]{
		\begin{tabular*}{\textwidth}{@{\extracolsep{\fill}} r c r r r}
			\toprule[0.12em]
			$\sqrt{s}$ [TeV] &  Inv.\ mass range [GeV] & $\sigma_{\text{signal}}$ [pb] & $\sigma_{\text{background}}$ [pb] & Ratio \\
			\midrule
			14 & 142 - 158 & $8.843 \cdot 10^{-2}$ & 2.921 & $3.027 \cdot 10^{-2}$ \\
			& 190 - 214 & $3.334 \cdot 10^{-2}$ & $9.938\cdot 10^{-1}$ & $3.356 \cdot 10^{-2}$ \\
			& 234 - 266 & $1.444 \cdot 10^{-2}$ & $5.111 \cdot 10^{-1}$ & $2.826 \cdot 10^{-2}$ \\
			& 378 - 426 & $2.149 \cdot 10^{-3}$ & $8.894 \cdot 10^{-2}$ & $2.416 \cdot 10^{-2}$ \\
			\midrule
			7 & 142 - 158 & $2.438 \cdot 10^{-2}$ & 1.228 & $1.986 \cdot 10^{-2}$ \\
			& 190 - 214 & $8.086 \cdot 10^{-3}$ & $4.078 \cdot 10^{-1}$ & $1.983 \cdot 10^{-2}$ \\
			& 234 - 266 & $3.159 \cdot 10^{-3}$ & $1.966 \cdot 10^{-1}$ & $1.607 \cdot 10^{-2}$ \\
			& 378 - 426 & $3.517 \cdot 10^{-4}$ & $3.045 \cdot 10^{-2}$ & $1.155 \cdot 10^{-2}$ \\
			\bottomrule[0.12em]
		\end{tabular*}
	\label{tbl:ComparisonResonancesBackgroundNoCuts}}\\
	\subfloat[After applying cuts on $\eta$ and $p_T$.]{
		\begin{tabular*}{\textwidth}{@{\extracolsep{\fill}} r c r r r}
			\toprule[0.12em]
			$\sqrt{s}$ [TeV] &  Inv.\ mass range [GeV] & $\sigma_{\text{signal}}$ [pb] & $\sigma_{\text{background}}$ [pb] & Ratio \\
			\midrule
			14 & 142 - 158 & $6.537 \cdot 10^{-2}$ & 1.318 & $4.959 \cdot 10^{-2}$ \\
			& 190 - 214 & $2.783 \cdot 10^{-2}$ & $5.931\cdot 10^{-1}$ & $4.692 \cdot 10^{-2}$ \\
			& 234 - 266 & $1.161 \cdot 10^{-2}$ & $2.665 \cdot 10^{-1}$ & $4.356 \cdot 10^{-2}$ \\
			& 378 - 426 & $1.589 \cdot 10^{-3}$ & $5.422 \cdot 10^{-2}$ & $2.931 \cdot 10^{-2}$ \\
			\midrule
			7 & 142 - 158 & $2.010 \cdot 10^{-2}$ & $6.771 \cdot 10^{-1}$ & $2.969 \cdot 10^{-2}$ \\
			& 190 - 214 & $7.439 \cdot 10^{-3}$ & $3.049 \cdot 10^{-1}$ & $2.440 \cdot 10^{-2}$ \\
			& 234 - 266 & $2.757 \cdot 10^{-3}$ & $1.301 \cdot 10^{-1}$ & $2.119 \cdot 10^{-2}$ \\
			& 378 - 426 & $3.218 \cdot 10^{-4}$ & $2.243 \cdot 10^{-2}$ & $1.435 \cdot 10^{-2}$ \\
			\bottomrule[0.12em]
		\end{tabular*}
	\label{tbl:ComparisonResonancesBackgroundCuts}}
	\caption{The MCPM cross-sections at the Higgs resonance peaks. The Higgs masses are 150, 200, 250 and 400 GeV, respectively.}
	\label{tbl:ComparisonResonancesBackground}
\end{table}

\cleardoublepage

\chapter*{Acknowledgements}

First and foremost, I would like to thank Dr.\ Markos Maniatis and Dr.\ Victor Lendermann for their very capable and ever-patient support during the course of this project. I would also like to express my sincere gratitude for my supervisors, Professor Hans-Christian Schultz-Coulon and Professor Otto Nachtmann, whose doors were always open for me. Finally, this thesis has benefitted greatly from discussions with Dominik Neuenfeld, Dr.\ Rainer Stamen, Merle Reinhart, Antje Brehmer, Jennifer Kieselmann, Eric Wisotzky and Dr.\ Martin Spiegel.

\addcontentsline{toc}{chapter}{Acknowledgements}

\cleardoublepage

\bibliography{References}

\begin{thebibliography}{30}
\providecommand{\natexlab}[1]{#1}
\providecommand{\url}[1]{\texttt{#1}}
\expandafter\ifx\csname urlstyle\endcsname\relax
  \providecommand{\doi}[1]{doi: #1}\else
  \providecommand{\doi}{doi: \begingroup \urlstyle{rm}\Url}\fi

\bibitem[Maniatis et~al.(2008{\natexlab{a}})Maniatis, von Manteuffel, and
  Nachtmann]{Maniatis_MCPMFamiliesMassHierarchy}
M.~Maniatis, A.~von Manteuffel, and O.~Nachtmann.
\newblock {A new type of CP symmetry, family replication and fermion mass
  hierarchies}.
\newblock \emph{Eur.Phys.J.}, C57:\penalty0 739--762, 2008{\natexlab{a}}.

\bibitem[{Particle Data Group}(2010)]{PDG}
{Particle Data Group}.
\newblock {Review of particle physics}.
\newblock \emph{J.Phys.G}, G37:\penalty0 075021, 2010.

\bibitem[{Nachtmann, O.}(1986)]{Nachtmann}
{Nachtmann, O.}
\newblock \emph{{Ph\"anomene und Konzepte der Elementarteilchenphysik}}.
\newblock Vieweg, 1986.

\bibitem[Maniatis et~al.(2008{\natexlab{b}})Maniatis, von Manteuffel, and
  Nachtmann]{Maniatis_CPViolationGeometry}
M.~Maniatis, A.~von Manteuffel, and O.~Nachtmann.
\newblock {CP violation in the general two-Higgs-doublet model: A Geometric
  view}.
\newblock \emph{Eur.Phys.J.}, C57:\penalty0 719--738, 2008{\natexlab{b}}.

\bibitem[Ferreira et~al.(2011)Ferreira, Haber, Maniatis, Nachtmann, and
  Silva]{Ferreira_THDMGeometry}
P.M. Ferreira, Howard~E. Haber, M.~Maniatis, O.~Nachtmann, and J.P. Silva.
\newblock {Geometric picture of generalized-CP and Higgs-family transformations
  in the two-Higgs-doublet model}.
\newblock \emph{Int.J.Mod.Phys.}, A26:\penalty0 769--808, 2011.

\bibitem[Maniatis et~al.(2006)Maniatis, von Manteuffel, Nachtmann, and
  Nagel]{Maniatis_THDMStability}
M.~Maniatis, A.~von Manteuffel, O.~Nachtmann, and F.~Nagel.
\newblock {Stability and symmetry breaking in the general two-Higgs-doublet
  model}.
\newblock \emph{Eur.Phys.J.}, C48:\penalty0 805--823, 2006.

\bibitem[Maniatis and Nachtmann(2009)]{Maniatis_MCPMPhenomenology}
M.~Maniatis and O.~Nachtmann.
\newblock {On the phenomenology of a two-Higgs-doublet model with maximal CP
  symmetry at the LHC}.
\newblock \emph{JHEP}, 0905:\penalty0 028, 2009.

\bibitem[Maniatis et~al.(2010)Maniatis, Nachtmann, and von
  Manteuffel]{Maniatis_MCPMPhenomenologyAddendum}
M.~Maniatis, O.~Nachtmann, and A.~von Manteuffel.
\newblock {On the phenomenology of a two-Higgs-doublet model with maximal CP
  symmetry at the LHC: Synopsis and addendum}.
\newblock 2010.

\bibitem[Maniatis and
  Nachtmann(2010)]{Maniatis_MCPMPhenomenologyRadiativeEffects}
M.~Maniatis and O.~Nachtmann.
\newblock {On the phenomenology of a two-Higgs-doublet model with maximal CP
  symmetry at the LHC. II. Radiative effects}.
\newblock \emph{JHEP}, 1004:\penalty0 027, 2010.

\bibitem[Maniatis and Nachtmann(2011)]{Maniatis_MCPMParameterSpace}
M.~Maniatis and O.~Nachtmann.
\newblock {Symmetries and renormalisation in two-Higgs-doublet models}.
\newblock \emph{JHEP}, 1111:\penalty0 151, 2011.

\bibitem[Alwall et~al.(2011)Alwall, Herquet, Maltoni, Mattelaer, and
  Stelzer]{MadGraph5}
J.~Alwall, M.~Herquet, F.~Maltoni, O.~Mattelaer, and T.~Stelzer.
\newblock {MadGraph 5: Going beyond}.
\newblock \emph{JHEP}, 1106:\penalty0 128, 2011.

\bibitem[Christenson et~al.(1964)Christenson, Cronin, Fitch, and
  Turlay]{CroninFitch_CPVIolation}
J.H. Christenson, J.W. Cronin, V.L. Fitch, and R.~Turlay.
\newblock {Evidence for the 2 pi decay of the K(2)0 Meson}.
\newblock \emph{Phys.Rev.Lett.}, 13:\penalty0 138--140, 1964.

\bibitem[Sakharov(1967)]{Sakharov_MatterAntimatter}
A.D. Sakharov.
\newblock {Violation of CP Invariance, C asymmetry, and baryon asymmetry of the
  universe}.
\newblock \emph{Pisma Zh.Eksp.Teor.Fiz.}, 5:\penalty0 32--35, 1967.

\bibitem[Peskin and Schroeder(1995)]{PeskinSchroeder}
M.E. Peskin and D.V. Schroeder.
\newblock \emph{An introduction to quantum field theory}.
\newblock Addison-Wesley Pub. Co., 1995.

\bibitem[Ecker et~al.(1987)Ecker, Grimus, and Neufeld]{Ecker_GeneralisedCP}
G.~Ecker, W.~Grimus, and H.~Neufeld.
\newblock {A standard form for generalized CP transformations}.
\newblock \emph{J.Phys.A}, A20:\penalty0 L807, 1987.

\bibitem[Georgi et~al.(1978)Georgi, Glashow, Machacek, and
  Nanopoulos]{Georgi_GluonFusion}
H.M. Georgi, S.L. Glashow, M.E. Machacek, and D.V. Nanopoulos.
\newblock {Higgs bosons from two gluon annihilation in proton proton
  collisions}.
\newblock \emph{Phys.Rev.Lett.}, 40:\penalty0 692, 1978.

\bibitem[Berger(2006)]{Berger}
C.~Berger.
\newblock \emph{Elementarteilchenphysik}.
\newblock Springer, 2006.

\bibitem[Berdine et~al.(2007)Berdine, Kauer, and Rainwater]{Berdine_NWA}
D.~Berdine, N.~Kauer, and D.~Rainwater.
\newblock {Breakdown of the narrow width approximation for new physics}.
\newblock \emph{Phys.Rev.Lett.}, 99:\penalty0 111601, 2007.

\bibitem[Alwall et~al.(2007)Alwall, Ballestrero, Bartalini, Belov, Boos,
  et~al.]{LHEF}
J.~Alwall, A.~Ballestrero, P.~Bartalini, S.~Belov, E.~Boos, et~al.
\newblock {A standard format for Les Houches event files}.
\newblock \emph{Comput.Phys.Commun.}, 176:\penalty0 300--304, 2007.

\bibitem[Sjostrand et~al.(2001)Sjostrand, Eden, Friberg, Lonnblad, Miu,
  et~al.]{PYTHIA}
T.~Sjostrand, P.~Eden, C.~Friberg, L.~Lonnblad, G.~Miu, et~al.
\newblock {High-energy physics event generation with PYTHIA 6.1}.
\newblock \emph{Comput.Phys.Commun.}, 135:\penalty0 238--259, 2001.

\bibitem[Asai(2006)]{GEANT}
M.~Asai.
\newblock {Geant4 --- a simulation toolkit}.
\newblock \emph{Trans.Amer.Nucl.Soc.}, 95:\penalty0 757, 2006.

\bibitem[Degrande et~al.(2011)Degrande, Duhr, Fuks, Grellscheid, Mattelaer,
  et~al.]{UFO}
C.~Degrande, C.~Duhr, B.~Fuks, D.~Grellscheid, O.~Mattelaer, et~al.
\newblock {UFO --- the universal FeynRules output}.
\newblock 2011.

\bibitem[Lai et~al.(2000)]{CTEQ5}
H.L. Lai et~al.
\newblock {Global QCD analysis of parton structure of the nucleon: CTEQ5 parton
  distributions}.
\newblock \emph{Eur.Phys.J.}, C12:\penalty0 375--392, 2000.

\bibitem[Pumplin et~al.(2002)Pumplin, Stump, Huston, Lai, Nadolsky,
  et~al.]{CTEQ6}
J.~Pumplin, D.R. Stump, J.~Huston, H.L. Lai, P.M. Nadolsky, et~al.
\newblock {New generation of parton distributions with uncertainties from
  global QCD analysis}.
\newblock \emph{JHEP}, 0207:\penalty0 012, 2002.

\bibitem[Brun and Rademakers(1997)]{ROOT}
R.~Brun and F.~Rademakers.
\newblock {ROOT: An object oriented data analysis framework}.
\newblock \emph{Nucl.Instrum.Meth.}, A389:\penalty0 81--86, 1997.

\bibitem[Various(2011)]{GNUPlot}
Various.
\newblock {GNUPlot documentation}.
\newblock \url{http://http://www.gnuplot.info/}, 2011.

\bibitem[{ATLAS collaboration}(1999)]{ATLAS_DesignReport}
{ATLAS collaboration}.
\newblock {ATLAS: Detector and physics performance technical design report.
  Volume 2}.
\newblock 1999.

\bibitem[{ATLAS collaboration}(2011{\natexlab{a}})]{ATLAS_Luminosity}
{ATLAS collaboration}.
\newblock {Luminosity public results}.
\newblock
  \url{https://twiki.cern.ch/twiki/bin/view/AtlasPublic/LuminosityPublicResults},
  2011{\natexlab{a}}.

\bibitem[Rizzo(2006)]{Rizzo_ZPrime}
T.G. Rizzo.
\newblock {$Z^\prime$ phenomenology and the LHC}.
\newblock pages 537--575, 2006.

\bibitem[{ATLAS collaboration}(2011{\natexlab{b}})]{ATLAS_ZPrimeSearch}
{ATLAS collaboration}.
\newblock {Search for dilepton resonances in pp collisions at $\sqrt{s} = 7$
  TeV with the ATLAS detector}.
\newblock \emph{Phys.Rev.Lett.}, 107:\penalty0 272002, 2011{\natexlab{b}}.

\end{thebibliography}

\end{document}